\documentclass[aps,pra,reprint,superscriptaddress]{revtex4-1}
\usepackage{amsmath} % Math typesetting, can't live without
\usepackage{mathtools} % Addition to amsmath
\usepackage{amsfonts} %
\usepackage{amssymb} %

\usepackage{color}
\usepackage{graphics}
\usepackage{epstopdf}
\usepackage[colorlinks=true,linkcolor=blue,citecolor=blue,urlcolor=blue]{hyperref} % Links/bookmarks for interactive PDFs

\usepackage{bbold} % Blackboard bold font, i.e. \mathbb{A} etc

\newcommand{\diff}{\mathrm{d}}
\newcommand{\transpose}{\mathrm{T}}
\newcommand{\trace}[2][]{\mathrm{Tr}_{#1}[#2]}
\newcommand{\diag}[1]{\mathrm{diag}\left(#1\right)}
\newcommand{\mrm}[1]{\mathrm{#1}}
\newcommand{\mbf}[1]{\mathbf{#1}}
\newcommand{\mcl}[1]{\mathcal{#1}}
\newcommand{\expo}[1]{\mathrm{e}^{#1}}

\newcommand{\sys}[1]{\mcl{#1}}
\newcommand{\op}[1]{#1}
\newcommand{\mat}[1]{\mbf{#1}}
\newcommand{\vect}[1]{\vec{#1}}

\begin{document}

%\preprint{}

\title{Open-system many-body dynamics through interferometric measurements and feedback}


%\author{Jonas Lammers}
%\email{jonas.lammers@itp.uni-hannover.de}
%\thanks{}
\author{Jonas Lammers}
\email{jonas.lammers@itp.uni-hannover.de}
\affiliation{Institute for Theoretical Physics, Leibniz Universit{\"a}t Hannover, Appelstra{\ss}e 2, 30167 Hannover, Germany}
\affiliation{Institute for Gravitational Physics (Albert Einstein Institute), Leibniz Universit{\"a}t Hannover, Callinstra{\ss}e 38, 30167 Hannover, Germany}
%\thanks{}
\author{Hendrik Weimer}
\affiliation{Institute for Theoretical Physics, Leibniz Universit{\"a}t Hannover, Appelstra{\ss}e 2, 30167 Hannover, Germany}
\author{Klemens Hammerer}
\affiliation{Institute for Theoretical Physics, Leibniz Universit{\"a}t Hannover, Appelstra{\ss}e 2, 30167 Hannover, Germany}
\affiliation{Institute for Gravitational Physics (Albert Einstein Institute), Leibniz Universit{\"a}t Hannover, Callinstra{\ss}e 38, 30167 Hannover, Germany}


\date{\today}

\begin{abstract}
Light-matter interfaces enable the generation of entangled states of light and matter which can be exploited to steer the quantum state of matter through measurement of light and feedback. Here we consider continuous-time, interferometric homodyne measurements of light on an array of light-matter interfaces followed by local feedback acting on each material system individually. While the systems are physically non-interacting, the feedback master equation we derive describes driven-dissipative, interacting many-body quantum dynamics, and comprises pairwise Hamiltonian interactions and collective jump operators. We characterize the general class of driven-dissipative many body systems which can be engineered in this way, and derive necessary conditions on models supporting non-trivial quantum dynamics beyond what can be generated by local operations and classical communication. We provide specific examples of models which allow for the creation of stationary many-particle entanglement, and the emulation of dissipative Ising models. Since the interaction between the systems is mediated via feedback only, there is no intrinsic limit on the range or geometry of the interaction, making the scheme quite versatile.
\end{abstract}

\pacs{}
%\keywords{}

\maketitle

\mathtoolsset{centercolon} % So the "equal define" sign ":=" looks right. Part of mathtools

\section{Introduction}
Great progress has been made in the construction of light-matter quantum interfaces between states of propagating photons and long-lived degrees of freedom of localized matter such as single ions \cite{Duan2010} and atoms \cite{Reiserer2015}, atomic ensembles \cite{Hammerer2004,Sangouard2011}, or equivalent solid state systems \cite{Schoelkopf2008}. A core functionality of such quantum interfaces is the efficient generation of entangled states among light and matter. These states, together with suitable interference and detection of photons, can be used to prepare entangled states of distant non-interacting quantum systems through entanglement swapping \cite{Cabrillo1999,Duan2001}. This has been demonstrated in a broad range of systems including pairs of single ions and atoms \cite{Moehring2007,Hofmann2012,Slodicka2013,Nolleke2013}, atomic ensembles \cite{Chou2005}, nitrogen-vacancy centers \cite{Bernien2013}, quantum dots \cite{Delteil2015}, rare-earth doped crystals \cite{Usmani2012}, and optical phonon modes in diamond \cite{Lee2011a}, see  \cite{Duan2010,Reiserer2015,Hammerer2004,Sangouard2011,Schoelkopf2008} for reviews. While the primary motivation for this development was to establish long-distance entanglement among quantum memory units for quantum communication \cite{Kimble2008}, recently, the same technique has enabled one of the first loophole-free violations of a Bell inequality \cite{Hensen2015}.

Looking further ahead, theoretical work by Barrett \textit{et al.} \cite{Barrett2005}, Lim \textit{et al.} \cite{Lim2005,Lim2006} and Vollbrecht and Cirac \cite{Vollbrecht2009} put forward another perspective: they showed that arrays of light-matter interfaces in combination with linear optical interferometry and subsequent detection of photons provides a resource for universal quantum computations as well as quantum simulations of Hamiltonian time evolutions of many-body systems \cite{Vollbrecht2009}. The key idea is that an $N$-port interferometric measurement of light can project an array of $N$ non-interacting material systems into correlated quantum states -- in the spirit of massive parallel entanglement swapping. A series of measurements, possibly including feedback operations depending on previous measurement results, effectively allows to execute quantum computations and simulations. These results clearly demonstrate the great potential offered by light-matter quantum interfaces in conjunction with interferometric measurements and feedback for the generation and manipulation of many-body quantum states.

In this article, we investigate the general class of setups where (i) light-matter interfaces emit continuous-wave light, (ii) light is mixed in a linear optical interferometer and (iii) measured in continuous homodyne detections, and (iv) the corresponding photocurrents are used to perform continuous-time feedback on the material systems. This is complementary to the schemes in \cite{Barrett2005,Lim2005,Lim2006,Vollbrecht2009} which have considered interfaces emitting single-photon pulses on demand in combination with photon counting measurements of interfering photons. Accordingly, quantum computations and simulations in \cite{Barrett2005,Lim2005,Lim2006,Vollbrecht2009} proceed probabilistically in discrete time steps, while the setup studied here gives rise to a deterministic continuous-time equation of motion for the steered material quantum systems.

We apply the powerful formalism for the description of quantum dynamics under continuous-time homodyne measurement and Markovian feedback introduced in \cite{Wiseman1993,Belavkin1994,Wiseman1994,Belavkin1995} and reviewed in \cite{Wiseman2010,Jacobs2014}. Our derivation, which is self-contained and does not require the use of stochastic calculus, reveals the general feedback master equation describing the corresponding open system many-body quantum dynamics. The feedback master equation exhibits effective Hamiltonians with pairwise interactions of arbitrary range and geometry along with collective jump operators composed of sums of strictly local operators. The setup thus allows to engineer the Hamiltonian and the dissipative part of system evolution, a topic which has received significant attention in recent years \cite{Muller2012,Daley2014}.
%%%%%%%%%%%%%%%%%%%%%%%%%%%%%%%%%%%%%%%%%%%%%%%%%%%%%%%%%%%%%%%%%%%%%%%%%%%%%%%%%%%%%%%%%%%%%%%%%%%%%%
%The simplest non-trivial setup of this type consists of two systems each coupled to a one-dimensional light field. In previous work we have shown that a suitable interference of the two light fields on a beam splitter and homodyne detection realizes a continuous-time Bell measurement on light \cite{Hofer2013}. This can be exploited to stabilize stationary entangled states of non-interacting material systems, as one of the authors showed in great detail for two optomechanical systems \cite{Hofer2015}. Here, we set out to generalize the original two-system--setup to an arbitrary number of systems, and to gauge its scope by attempting to reach two complementary objectives. First, we aim for the deterministic generation of a multipartite entangled state in the spirit of \cite{Hofer2013,Hofer2015}. Second, motivated by the growing interest in quantum simulators, we attempt to realize a given quantum dynamics. As a concrete example, we generate an Ising Hamiltonian with a transverse field, which is accompanied by collective jump operators inherent to our approach.

The simplest non-trivial setup of this type consists of two systems each coupled to a one-dimensional light field. In previous work we have shown that a suitable interference of the two light fields on a beam splitter and homodyne detection realizes a continuous-time Bell measurement on light \cite{Hofer2013}. This can be exploited to stabilize stationary entangled states of non-interacting material systems, as one of the authors showed in great detail for two optomechanical systems \cite{Hofer2015}.

We would like to highlight three features of this approach: First, entanglement is generated deterministically, independent of the initial state, through properly designed dissipation similar to schemes presented in, e.g., \cite{Barreiro2010,Krauter2011,*Muschik2011,Stannigel2012}. We emphasize that this is different from strategies to preserve pre-existing entanglement, see \cite{Suter2016} for a review, as well as \cite{Man2015} and references therein. Second, the collective dissipation and interaction is realized entirely through measurement and feedback. It does not require direct or coherent coupling of the systems, unlike \cite{Barreiro2010,Krauter2011,*Muschik2011,Stannigel2012}. Third, the scheme is naturally scalable to a many-body setup akin to those proposed in \cite{Diehl2008,Kraus2008,Verstraete2009}.

Here, we set out to generalize the original two-system--setup considered in \cite{Hofer2013,Hofer2015} to an arbitrary number of systems, and to gauge its scope by attempting to reach two complementary objectives. First, we aim for the deterministic generation of a multipartite entangled state in the spirit of \cite{Hofer2013,Hofer2015}. Second, motivated by the growing interest in quantum simulators, we attempt to realize a given quantum dynamics. As a concrete example, we generate an Ising Hamiltonian with a transverse field, which is accompanied by collective jump operators inherent to our approach.
%%%%%%%%%%%%%%%%%%%%%%%%%%%%%%%%%%%%%%%%%%%%%%%%%%%%%%%%%%%%%%%%%%%%%%%%%%%%%%%%%%%%%%%%%%%%%%%%%%%%%%

The paper is organized as follows. In Section \ref{sec:Derivation} we present the derivation of the Feedback Master Equation (FME) governing the collective system dynamics. Section \ref{sec:LOCC} then discusses non-trivial conditions leading to \textit{local operations and classical communication (LOCC)} dynamics precluding true quantum dynamics. Applications are considered in the succeeding sections with an $N$-qubit generalization of a 2-qubit protocol \cite{Hofer2013,Hofer2015} in Section \ref{sec:StatePreparation}, and by engineering a dissipative Ising model in Section \ref{sec:QuantumSimulation}. Conclusion and outlook follow in Section \ref{sec:Conclusion}. 
\section{\label{sec:Derivation}Derivation of the feedback master equation}

\begin{figure}
	\includegraphics[width=.98\columnwidth,keepaspectratio=true]{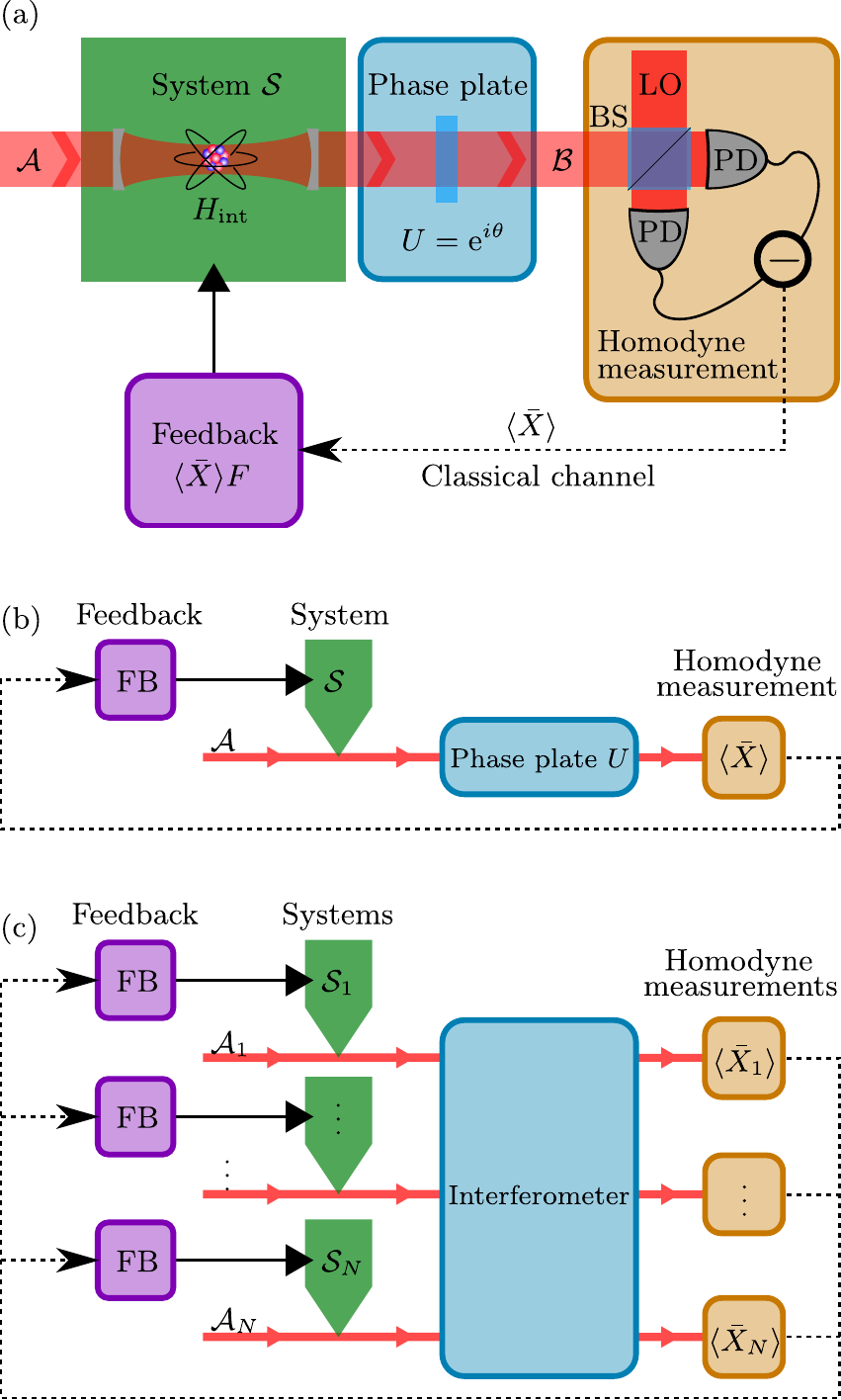}%
	\caption{(a) Feedback setup for a single system $\sys{S}$, such as an atom in a cavity (schematic). A one-dimensional light field $\sys{A}$ couples to $\sys{S}$ via $\op{H}_\mrm{int}$, and passes through a phase plate $U=\expo{i\theta}$. Subsequently, the transformed field $\sys{B}=U\sys{A}$ is combined with a strong local oscillator (LO) on a beam splitter (BS). Two photodetectors (PD) measure the incident intensity. The difference of the photocurrents then yields the desired quadrature $\langle \op{\bar{X}} \rangle$. This classical signal is transmitted back to the system (dotted line), where it is used to generate Hermitian feedback $\langle \op{\bar{X}} \rangle\op{F}$. (b) Schematic representation of the previous setup, using the same color code. (c) Proposed generalization to $N$ systems $\sys{S}_{k}$, $k=1,\dots,N$, with corresponding light fields $\sys{A}_{k}$. After interacting with the systems, the fields traverse an $N$-port passive interferometer. Homodyne measurement of each beam yields quadratures $\langle \op{\bar{X}}_{k} \rangle$, which together allow us to steer the systems via local feedback. \label{fig:Derivation:SingleSystemFeedback}}
\end{figure}

The description of quantum dynamics under continuous diffusive measurement (that is, in our context, homodyne detection of light) and feedback is thoroughly described in the formalism of stochastic Schr\"odinger and master equations, as pioneered by Belavkin \cite{Belavkin1992} and Parthasarathy \cite{Hudson1984}, and summarized comprehensively in the recent textbooks of Wiseman and Milburn \cite{Wiseman2010} and Jacobs \cite{Jacobs2014}. This formalism provides stochastic equations of motion for the quantum states \textit{conditioned} on the measurement result (that is, the photocurrent). The equation of motion for the \textit{unconditional} quantum state can then easily be obtained in the ensemble average over all measurement results, which removes all stochastic terms. Following \cite{Wiseman2010}, we refer to the equation of motion describing unconditional, ensemble averaged quantum dynamics under continuous measurement and feedback as the Feedback Master Equation (FME). In this section we provide a derivation of the Feedback Master Equation (FME) which avoids the use of stochastic calculus, and is based solely on the formalism of basic quantum mechanics. A similar presentation has been given before by Hofer \textit{et al.}\ \cite{Hofer2013,Hofer2015} for the special case of one or two monitored systems, and it is our main aim here to extend this approach to setups involving many systems. In the following, we first recapitulate the case of a single system coupled to a single light field in detail, cf. Fig.~\hyperref[fig:Derivation:SingleSystemFeedback]{\ref{fig:Derivation:SingleSystemFeedback}~(a,b)}. The generalization to $N$ systems coupled to $M\geq N$ beams of light and feedback fields is then straightforward, cf. Fig.~\hyperref[fig:Derivation:SingleSystemFeedback]{\ref{fig:Derivation:SingleSystemFeedback}~(c)}.

\subsection{Feedback master equation of a single system}
We consider a generic physical setup as shown in Fig.~\hyperref[fig:Derivation:SingleSystemFeedback]{\ref{fig:Derivation:SingleSystemFeedback}~(a)} which is composed of a system $\sys{S}$ (e.g., a mode of a cavity or a single atom, enclosed by the light green box) that is coupled to a one-dimensional light field $\sys{A}$ (e.g., a paraxial beam or a field in a waveguide). The combined evolution of $\sys{S}$ and $\sys{A}$ is governed by a general Hamiltonian $\op{H} = \op{H}_{\sys{S}} + \op{H}_{\sys{A}} + \op{H}_{\mrm{int}}$, comprising Hamiltonians for the system, the light field, and their interaction, respectively. The Hamiltonian of the free field reads	$\op{H}_{\sys{A}} = \int_{0}^{\infty} \diff \omega\, \omega \hat{a}_{\omega}^{\dag} \hat{a}_{\omega}$, where $\hat{a}_{\omega}$ is the annihilation operator of the mode with frequency $\omega$ obeying $[\hat{a}_\omega,\hat{a}^\dagger_{\omega'}]=\delta(\omega-\omega')$. The interaction is assumed to be of the generic form
\begin{equation}
	\op{H}_{\mrm{int}} = i\int_0^\infty\diff\omega \sqrt{\frac{\kappa(\omega)}{2\pi}} \left( \op{S}\hat{a}^{\dag}_{\omega} - \op{S}^\dag \hat{a}_{\omega} \right),% \label{eq:Derivation:InteractionHamiltonian}
\end{equation}
with a system operator $\op{S}$ and coupling strength $\kappa(\omega)$.

We go to an interaction frame with respect to $\op{H}_{\sys{S}}+\op{H}_{\sys{A}}$, and make the following assumptions. Firstly, the system-light interaction is constant in some frequency bandwidth $\mcl{W}$, that is $\kappa(\omega)\equiv \kappa$ for $\omega\in\mcl{W}$; Secondly, the system operators merely acquire a dominant time dependent phase $\op{S}(t)=\op{S} \expo{-i\Omega t}$ in the interaction picture. Any further time dependence is negligible on the time scale of the interaction $\tau_\mathrm{int}\simeq \kappa^{-1}$. In the rotating frame and using these assumptions we write the interaction as
\begin{equation}
	\op{H}_{\mrm{int}}(t) = i\sqrt{\kappa}\left( \op{S}\hat{a}^{\dag}(t) - \op{S}^\dag \hat{a}(t) \right), \label{eq:Derivation:InteractionHamiltonian}
\end{equation}
with time-independent system operators $\op{S}$, and time-dependent field operators $\hat{a}(t):=(2\pi)^{-1/2} \int_\mcl{W} \diff\omega\, \hat{a}_{\omega} \expo{-i(\omega-\Omega) t}$.
%
%Note that the field operators $\hat{a}(t)$ satisfy the commutation relation $[\hat{a}(t),\hat{a}^\dag(t')] = \delta(t-t')$. Intuitively, this is because the interaction with $\sys{S}$ is localized at some point $z_0$ along the beam-axis of $\sys{A}$, and we implicitly consider only operators $\op{a}(t)\equiv \op{a}(t,z_0)$ at this location. The part of the light field which is at $z_0$ at time $t$ has no way to interact with a part of the field which reaches $z_0$ at some other time $t'$.

To illustrate this, we consider a few examples. First, let the system be a single mode cavity with annihilation operator $\hat{c}$. In that case, coupling to the outside field is given by $\op{S}\propto \hat{c}$. If the system is a simple two-level atom with raising and lowering operators $\hat{\sigma}_{\pm}$, the usual dipole coupling is given by $\op{S}\propto \hat{\sigma}_{+} + \hat{\sigma}_{-}$, which may simplify to $\op{S}\propto \hat{\sigma}_{-}$ in the rotating wave approximation \cite{Wiseman2010}.
An (effective) spin-$\frac{1}{2}$ particle, such as a two-level atom inside a far-detuned cavity with a linearized field \cite{Thomsen2002}, may correspond to $\op{S}\propto \hat{\sigma}_{z}$ with the usual Pauli operator $\hat{\sigma}_{z}$.
% A spin-$\frac{1}{2}$ particle inside a far-detuned cavity will, with a linearized field, correspond to $\op{S}\propto \hat{\sigma}_{z}$ with Pauli operator $\hat{\sigma}_{z}$ \cite{Thomsen2002}.
Lastly, coupling a mechanical oscillator with quadrature $\hat{x}$ to light, and linearizing the interaction and field, yields a radiation-pressure-interaction $\op{S}\propto \hat{x}$ \cite{Aspelmeyer2014}. We emphasize that the assumptions leading to Eq. \eqref{eq:Derivation:InteractionHamiltonian} hold in a broad range of systems, making the following considerations applicable to a variety of physical realizations.

\subsubsection{Coarse-graining of time\label{sec:Derivation:CoarseGrainingOfTime}}
Let us now consider how the combined system-light state $\rho^{\sys{S}\sys{A}}(t)$ evolves (in the interaction frame) during a small time step $t\to t+\delta t$. % The step $\delta t\sim \tau_\mrm{int}$ is initially chosen on the order of the interaction time.
The evolution is then governed by an interaction of $\sys{S}$ with a discrete spatial part of $\sys{A}$ of length $c\,\delta t$, where $c$ is the speed of light. Coarse-graining time in this way is justified since any physical measurement will have limited temporal resolution. Assuming a sufficiently strong and fast interaction allows us to eventually consider the limit of infinitesimal $\delta t\to\diff t$ to derive the continuous evolution of $\sys{S}$.

Since at each step in time a different part of the light field moves in, it is justified to assume that $\sys{A}$ and $\sys{S}$ initially form a product state,
\begin{align}
	\rho^{\sys{S}\sys{A}}(t) = \rho^{\sys{S}}(t)\otimes\rho^{\sys{A}},	
\end{align}
and that the incoming field is in a vacuum state, $\rho^{\sys{A}}=|\mrm{vac}\rangle \langle \mrm{vac}|$. Interaction during the time $\delta t$ then generates the state
\begin{align}
	\rho_\mrm{int}^{\sys{S}\sys{A}} := U(t+\delta t, t)\rho^{\sys{S}\sys{A}}(t)U^\dag (t+\delta t, t).
\end{align}
Here $U$ is the usual time-ordered evolution operator,
\begin{align}
	U(t+\delta t, t) & := \mcl{T}\exp\left(-i\int_{t}^{t+\delta t}\diff t'\, \op{H}_\mrm{int}(t')\right),
\end{align}
which may be expanded as (see, e.g., \cite[Chapter 4.2]{Peskin1995})
\begin{align}
	U(t+\delta t, t) & = 1 + U^{(1)} + U^{(2)} + \mcl{O}(\delta t^{3/2}) \label{eq:Derivation:UnitaryExpansion}
\end{align}
with (non-unitary) operators
\begin{align}
	U^{(1)} & = - i\int_{t}^{t+\delta t}\diff t_1\, \op{H}_\mrm{int}(t_1), \tag{\ref*{eq:Derivation:UnitaryExpansion}a}\\
	U^{(2)} & = - \int_{t}^{t+\delta t}\diff t_1\,\int_{t}^{t_1}\diff t_2\, \op{H}_\mrm{int}(t_1)\op{H}_\mrm{int}(t_2). \tag{\ref*{eq:Derivation:UnitaryExpansion}b}
\end{align}
Due to the singular nature of $\hat{a}(t)$ and $\hat{a}^\dag(t)$, we have to consider the first two components of the series to capture all terms of order $\delta t$ \footnote{Intuitively, treating $\delta t$ as the smallest possible time increment, each term of the expansion goes as a power of $\sqrt{\delta t}$: each factor of $\hat{a}$ goes as $\delta t^{-1/2}$ (since $[\hat{a}(t),\hat{a}^\dag (t)]=\delta(0)\sim 1/\delta t$ \cite{Wiseman2010}) while each integration $\int^{\delta t}\diff t'\dots$ contributes a factor of $\delta t$.}.
We rewrite the integrand of $U^{(2)}$ using $\op{H}_\mrm{int}(t_1)\op{H}_\mrm{int}(t_2) = \op{H}_\mrm{int}(t_1)\op{H}_\mrm{int}(t_2)/2 + \op{H}_\mrm{int}(t_2)\op{H}_\mrm{int}(t_1)/2 + [ \op{H}_\mrm{int}(t_1),\op{H}_\mrm{int}(t_2) ]/2$, and we find that the commutator vanishes on the vacuum state. Being left with a symmetric integrand we may change the integration boundaries to obtain
\begin{align}
	U^{(2)}|\mrm{vac}\rangle & = - \frac{1}{2}\int_{t}^{t+\delta t}\diff t_1\,\int_{t}^{t+\delta t}\diff t_2\, \notag\\
	& \qquad\qquad\qquad\quad \times \op{H}_\mrm{int}(t_1)\op{H}_\mrm{int}(t_2)|\mrm{vac}\rangle.
\end{align}
Formally defining coarse-grained, dimensionless operators
\begin{subequations}\label{eq:Derivation:CoarseGraining}
	\begin{align}
		\op{\bar{A}} & := \frac{1}{\sqrt{\delta t}}\int_{t}^{t+\delta t}\diff t'\, \hat{a}(t'), \\
		\op{\bar{H}}_\mrm{int} & := i\left(S\bar{A}^\dag - \op{S}^\dag \op{\bar{A}} \right),
	\end{align}
\end{subequations}
we can write the effect of $\op{U}$ on the vacuum as
\begin{align}
	U(t+\delta t, t)|\mrm{vac}\rangle \approx \exp\left( -i\epsilon\op{\bar{H}}_\mrm{int} \right)|\mrm{vac}\rangle, \label{eq:Derivation:CoarseGrainedEvolutionOperator}
\end{align}
where $\epsilon=\sqrt{\kappa\delta t}$ and the identity in \eqref{eq:Derivation:CoarseGrainedEvolutionOperator} holds if we neglect terms of order $\mcl{O}(\epsilon^3)$. The evolved system-light state, also normalized up to order $\mcl{O}(\epsilon^{3})$, is thus given by
\begin{align}
	\rho_\mrm{int}^{\sys{S}\sys{A}} \approx \exp\left( -i\epsilon\op{\bar{H}}_\mrm{int} \right) \rho^{\sys{S}\sys{A}}(t) \exp\left( i\epsilon\op{\bar{H}}_\mrm{int} \right). \label{eq:Derivation:InteractionDensMat}
\end{align}

Note that the normalization of $\bar{A}$ in Eq.~\eqref{eq:Derivation:CoarseGraining} was chosen to realize the commutation relation
\begin{align}
	[\bar{A},\bar{A}^\dag] & = \frac{1}{\delta t}\int\limits_{t}^{t+\delta t} \diff t'\int\limits_{t}^{t+\delta t}\diff t'' [\hat{a}(t'),\hat{a}^\dag(t'')]  =  1,
\end{align}
independent of the measurement time $\delta t$. %While the original time-ordered exponential implicitly depended on $\delta t$, this dependence has now been extracted and made apparent.
The commutation relation shows that we may treat $\bar{A}$ as the annihilation operator of a single bosonic mode corresponding to a temporal ``slice'' of the incoming field of duration $\delta t$.

\subsubsection{Phase-plate and measurement}
After it has interacted with the system $\sys{S}$, we let the field $\sys{A}$ traverse a passive optical element, which for a single one-dimensional field corresponds simply to a phase plate (light blue box in Fig.~\hyperref[fig:Derivation:SingleSystemFeedback]{\ref{fig:Derivation:SingleSystemFeedback}~(a)}). In later sections we will consider setups involving several fields in which case the passive element corresponds to a linear interferometer.

In order to pave the way for this generalization we give here an overly elaborate description of the action of a phase plate. We assume the phase plate to be independent of frequency in the relevant bandwidth. The plate imprints an additional phase $\theta$ on $\sys{A}$, so $\bar{A}\mapsto U\op{\bar{A}} := \expo{i\theta}\bar{A}$. We view this phase-shifted field as a new field $\sys{B}$, described by the field operator $\op{\bar{B}} := U\bar{A}$, which still satisfies $[\bar{B},\bar{B}^{\dag}]=1$. Simultaneously defining a new system operator
\begin{align}
	\op{Z} := U\op{S} = \expo{i\theta} \op{S}
\end{align}
allows us to write $\op{\bar{H}}_\mrm{int} = i\left( Z\bar{B}^\dag - \op{Z}^\dag\bar{B}\right)$. % While these redefinitions seem redundant in the case of a single system, they later allow for straightforward generalization of the formalism to multiple systems.

The next step in our scheme is a homodyne quadrature measurement of the outgoing beam $\sys{B}$, enclosed in the dark orange box in Fig.~\hyperref[fig:Derivation:SingleSystemFeedback]{\ref{fig:Derivation:SingleSystemFeedback}~(a)}. The quadrature operators of $\sys{B}$ are given by $\op{\bar{X}} := \left(\bar{B}+\bar{B}^\dag\right)/\sqrt{2}$, and $\bar{P} := -i\left(\bar{B}-\bar{B}^\dag\right)/\sqrt{2}$, with commutation relation $[\op{\bar{X}},\bar{P}]=i$. Without loss of generality we restrict ourselves to measuring only $\op{\bar{X}}$, since we can always change the quadrature by choosing a different phase $\theta$. The $\op{\bar{X}}$-quadrature eigenstates $|\bar{x}\rangle$ of the field $\sys{B}$ are defined through $\op{\bar{X}}|\bar{x}\rangle = \bar{x}|\bar{x}\rangle$ for $\bar{x}\in\mathbb{R}$.

Assuming we have measured a specific value, $\bar{x}$ say, the conditional state of $\sys{S}$ after interaction and projective light measurement is then given by
\begin{align}
	\rho_\mrm{cond}^{\sys{S}}(\bar{x}) := \frac{1}{p(\bar{x})}\langle \bar{x}| \rho_\mrm{int}^{\sys{S}\sys{A}}|\bar{x} \rangle, \label{eq:Derivation:ConditionalState}
\end{align}
with normalization $p(\bar{x})=\trace[\sys{S}]{\langle \bar{x}|\rho_\mrm{int}^{\sys{S}\sys{A}} |\bar{x} \rangle}=\trace{|\bar{x}\rangle\langle \bar{x}| \rho_\mrm{int}^{\sys{S}\sys{A}}}$. Here, $\trace[\sys{S}]{\dots}$ denotes a partial trace over the system.

\subsubsection{Applying feedback}
Using the measurement result $\bar{x}$, we apply instantaneous Markovian feedback to $\rho_\mrm{cond}^{\sys{S}}(\bar{x})$, depicted by the purple box in Fig.~\hyperref[fig:Derivation:SingleSystemFeedback]{\ref{fig:Derivation:SingleSystemFeedback}~(a)}. \textit{Instantaneous} here means that all possible delay is negligible on the time-scale $\delta t$, so that it can act back on the system before the succeeding time-step. Physically this treatment is justified if the delay is smaller than the interaction time $\tau_\mrm{int}$. The feedback is \textit{Markovian} in the sense that it depends only on the measured signal at a specific point in time, and not on a record of previous measurements.

We aim to keep the feedback simple to alleviate experimental difficulties. To this end we choose \textit{Hamiltonian feedback} \cite{Wiseman1993,Wiseman1994} (also known as \textit{Wiseman-Milburn Markovian feedback} \cite{Jacobs2014}), which is linear in the measurement result $\bar{x}$. It is effected by modifying the Hamiltonian proportional to the unprocessed measurement signal, and would enter a stochastic master equation as $[\dot{\rho}]_\text{fb} = -i[\op{F},\rho]I(t)$, where $I(t)\propto \bar{x}(t)$ is the measured photocurrent and $\op{F}$ is some Hermitian system operator.

For example, let the system be a single mode cavity with annihilation (creation) operator $\hat{c}$ $(\hat{c}^{\dag})$. Then $\op{F} \propto \hat{c}^\dag\hat{c}$ shifts the frequency of the cavity mode \cite{Horoshko1997}, whereas $\op{F} \propto i(\hat{c}-\hat{c}^\dag)$ affects the cavity drive \cite{Wiseman1994a}. Physically, these operations may be realized by modulating either the refractive index inside the cavity or its transmission.
On the other hand, if the system is a particle with (effective) spin-$\frac{1}{2}$, we may rotate its spin through $\op{F}\propto \hat{\sigma}_{y}$ with Pauli operator $\hat{\sigma}_{y}$. This is realized, for example, for a two-level atom inside a cavity by modulating the amplitude of an applied radio-frequency magnetic field \cite{Thomsen2002}.
%On the other hand, if the system is a two-level atom, we may rotate its (effective) spin by modulating the amplitude of a radio-frequency magnetic field. This effects $\op{F}\propto \hat{\sigma}_{y}$ \cite{Thomsen2002}.
Further examples of feedback operators are provided in the applications in Sections \ref{sec:StatePreparation} and \ref{sec:QuantumSimulation} (see, for instance, Eqs. \eqref{eq:StatePreparation:2QP:SingleFeedback}, \eqref{eq:StatePreparation:2QP:MultiFeedback}, and \eqref{eq:QuantumSimulation:Feedback}).

In our coarse-grained description the system state after feedback is given by
\begin{align}
	\rho_\mrm{fb}^{\sys{S}}(\bar{x}) & := \expo{-i\sqrt{2}\epsilon\bar{x}F}\rho_\mrm{cond}^{\sys{S}}(\bar{x})\expo{i\sqrt{2}\epsilon\bar{x}F}. \label{eq:Derivation:FeedbackState}
\end{align}
The factor $\epsilon$ appears in the exponent as the natural scaling of $\op{\bar{X}}$ due to the coarse-graining in Eq.~\eqref{eq:Derivation:CoarseGraining}, and the arbitrary factor $\sqrt{2}$ was chosen to obtain a more convenient master equation eventually. Note that this type of feedback does not require any quantum information to be transmitted back to the system, only the measurement result.

Let us express the conditional system state $\rho_\mrm{fb}^{\sys{S}}(\bar{x})$ in terms of the initial states of $\sys{S}$ and $\sys{A}$. Plugging in the definitions of $\rho_\mrm{cond}^{\sys{S}}$ and $\rho_\mrm{int}^{\sys{S}\sys{A}}$ from Eqs. \eqref{eq:Derivation:ConditionalState} and \eqref{eq:Derivation:FeedbackState}, and using the fact that $|\bar{x}\rangle$ is an eigenstate of $\op{\bar{X}}$, we find
\begin{align}
	\rho_\mrm{fb}^{\sys{S}}(\bar{x}) & = \frac{1}{p(\bar{x})} \langle \bar{x}| \expo{-i\sqrt{2}\epsilon\op{\bar{X}}\op{F}} \rho_\mrm{int}^{\sys{S}\sys{A}} \expo{i\sqrt{2}\epsilon\op{\bar{X}}\op{F}} |\bar{x} \rangle \notag\\
	& = \frac{1}{p(\bar{x})} \langle \bar{x}| \op{K}\left(\rho^\sys{S}\otimes \rho^{\sys{A}}\right) \op{K}^\dagger |\bar{x} \rangle, \label{eq:Derivation:ConditionalFeedbackState}
\end{align}
where
\begin{align}
	K:=\exp({-i\sqrt{2}\epsilon\op{\bar{X}}\op{F}}) \exp({-i\epsilon\op{\bar{H}}_\mrm{int}}).
\end{align}

\subsubsection{Unconditional master equation}
By definition, the feedback, and thus $\rho_\mrm{fb}^{\sys{S}}(\bar{x})$, depends on a particular measurement result. To obtain the \textit{unconditional} evolution of $\sys{S}$ we have to consider and average over all possible results $\bar{x}$, weighted with their respective probability $p(\bar{x})=\trace{|\bar{x}\rangle \langle \bar{x}|\rho_\mrm{int}^{\sys{S}\sys{A}}}$. With $\rho_\mrm{fb}^{\sys{S}}(\bar{x})$ from Eq.~\eqref{eq:Derivation:ConditionalFeedbackState} this average is
\begin{align}
	\rho^{\sys{S}}(t+\delta t) & = \int_{-\infty}^{\infty} \diff \bar{x}\, p(\bar{x})\rho_\mrm{fb}^{\sys{S}}(\bar{x})
	\notag\\
	& = \trace[\sys{A}]{\op{K}\left(\rho^\sys{S}(t)\otimes |\mrm{vac}\rangle\langle\mrm{vac}|\right) \op{K}^\dagger},
\end{align}
where the trace $ \trace[\sys{A}]{\dots} = \int\diff \bar{x}\, \langle \bar{x}|\dots |\bar{x} \rangle$ denotes a partial trace over the light field.

We evaluate the trace by expanding $\op{K}$ in powers of $\epsilon$ and neglecting terms of order $\mcl{O}(\epsilon^{3})$. This yields
\begin{align}
	\rho^\sys{S}(t+\delta t) & \approx \rho^{\sys{S}}(t)-\frac{\epsilon^2}{2}\left\{ (\op{Z}^\dag Z+\op{Z}^2)\rho^{\sys{S}}(t) + \mrm{h.c.} \right\} \notag\\
	& \quad + \sqrt{2}\epsilon\langle \bar{x} \rangle\left\{ (\op{Z}-i\op{F})\rho^{\sys{S}}(t) +\mrm{h.c.} \right\} \notag\\
	& \quad + \epsilon^2\langle \bar{x}^2 \rangle\left\{ (\op{Z}^2 -\op{F}^2-2i\op{F}\op{Z})\rho^{\sys{S}}(t) + \mrm{h.c.}\right\} \notag\\
	& \quad + 2\epsilon^2 \langle \bar{x}^2 \rangle(\op{Z}-i\op{F})\rho^{\sys{S}}(t)(\op{Z}^\dag +iF),
\end{align}
with system operators $\op{F}$ and $\op{Z}=U\op{S}=\expo{i\theta}\op{S}$.
To simplify the notation, let us drop the system superscript, $\rho^{\sys{S}}\equiv\rho$. We can use the vacuum statistics of $\bar{x}$ \cite{Wiseman2010}, namely $\langle \bar{x} \rangle = 0$ and $\langle \bar{x}^2 \rangle = 1/2$, and $\epsilon=\sqrt{\kappa\delta t}$ to obtain the unconditional system state
\begin{align}\label{eq:Derivation:ExplicitSystemEvolution}
	\rho(t+\delta t) \approx \rho(t) & -\frac{i}{2}[\op{F}\op{Z}+\op{Z}^\dag \op{F},\rho(t)] \kappa\delta t \notag\\
	& + \mcl{D}[\op{Z}-i\op{F}]\rho(t) \kappa\delta t,
\end{align}
with the usual Lindblad superoperator $\mcl{D}$ defined by
\begin{align}
	\mcl{D}[J]\rho := J\rho J^\dag - \frac{1}{2}(J^\dag J\rho + \rho J^\dag J).
\end{align}
From Eq.~\eqref{eq:Derivation:ExplicitSystemEvolution} it is clear that the time scale of the corresponding quantum dynamics is set by the strength of the light-matter interaction $\kappa$. In the following we will suppress this overall scaling by setting $\kappa=1$ which corresponds either to absorbing $\kappa$ into the definitions of the system and feedback operators $\op{S}$ and $\op{F}$, or equivalently, by interpreting $t$ as a dimensionless time measured in units of $\tau_\mathrm{int}=\kappa^{-1}$.

Let us now consider the limit $\delta t \to \diff t$ of an infinitesimal step in the evolution. This is straightforward since all of the operators appearing in Eq.~\eqref{eq:Derivation:ExplicitSystemEvolution} are independent of time, so we find the infinitesimal state increment
\begin{align}
	\diff \rho & = \rho(t+\diff t) - \rho(t)\notag\\
	& = -\frac{i}{2}[\op{F}\op{Z}+\op{Z}^\dag \op{F},\rho]\diff t + \mcl{D}[\op{Z}-i\op{F}]\rho\diff t.
\end{align}
We can read off the Feedback Master Equation (FME)
\begin{align}
	\dot{\rho} = -i[\op{H},\rho] + \mcl{D}[\op{J}]\rho \label{eq:Derivation:1DFME}
\end{align}
with Hamiltonian $\op{H}$ and jump operator $\op{J}$ respectively given by
\begin{align}
	\op{H} & := \frac{1}{2}(\op{F}\op{Z}+\op{Z}^\dag \op{F}), \tag{\ref*{eq:Derivation:1DFME}a}\\
	\op{J} & := \op{Z}-i\op{F}. \tag{\ref*{eq:Derivation:1DFME}b}
\end{align}
We arrived at this important result using the basic rules of quantum mechanics, but emphasize that one obtains the same result \cite{Hofer2013,Hofer2015} by treating the measurement and feedback using stochastic calculus \cite{Belavkin1992,Wiseman2010}. We also point out that the equation of motion of the conditional state (without performing the ensemble average) can be formulated only using stochastic calculus.

\subsection{Generalization to multiple systems\label{sec:Derivation:Generalization}}
\begin{figure}
	\includegraphics[width=\columnwidth,keepaspectratio=true]{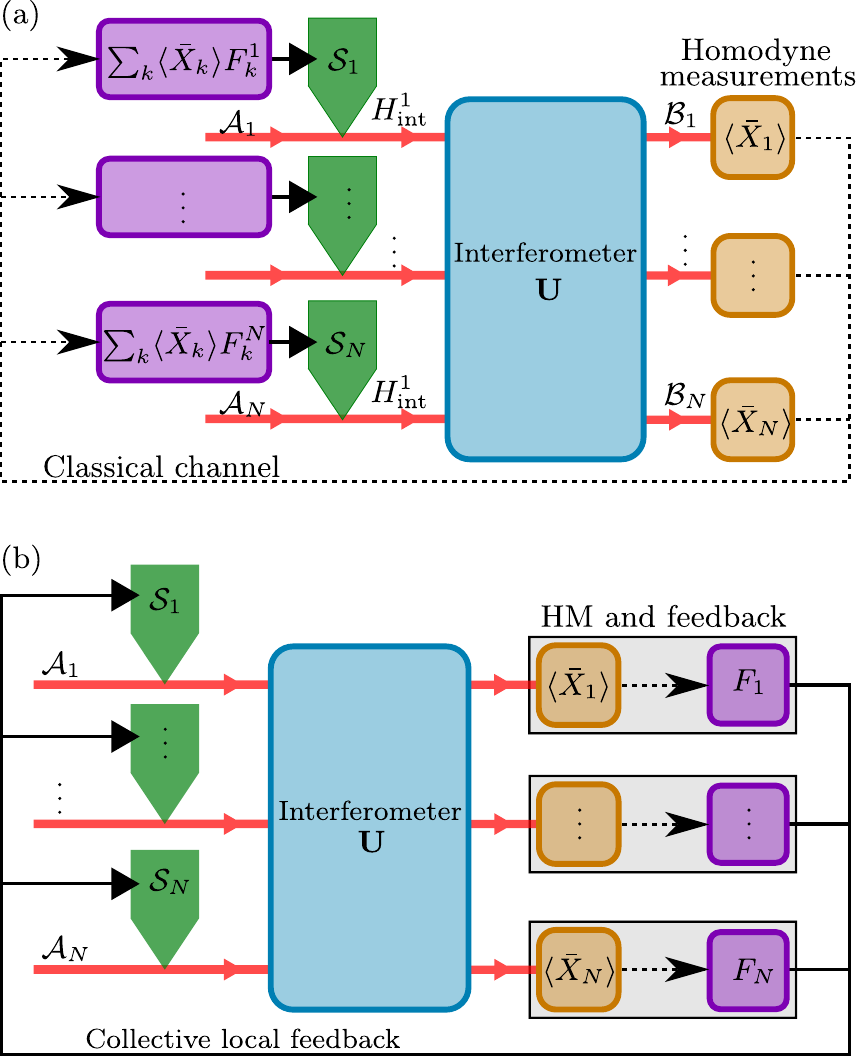}%
	\caption{(a) Schematic layout of the feedback scheme for $N$ systems $\sys{S}_{1}$ through $\sys{S}_{N}$. Each system interacts with a single light field $\sys{A}_{k}$ via $\op{H}_{\mrm{int}}^{k}$. The light then traverses an interferometer given by the $N\times N$ unitary matrix $\mat{U}$. We perform a homodyne measurement of each resulting field $\sys{B}_{k}$, which yields the quadratures $\langle \op{\bar{X}}_{k} \rangle$. Each signal is transmitted back to the systems via classical channels (dotted line), and generates local feedback $\langle \op{\bar{X}}_k \rangle \op{F}_{k}^{l}$, where $\op{F}_{k}^{l}$ acts only on $\sys{S}_{l}$. The total feedback on $\sys{S}_{l}$ is generated by $\sum\nolimits_{k=1}^{N} \langle \op{\bar{X}}_{k} \rangle \op{F}_{k}^{l}$. (b) A different schematic of the same setup. It emphasizes how each measurement $\langle \op{\bar{X}}_k \rangle$ is used to generate collective feedback $\op{F}_k = \sum\nolimits_{l=1}^{N}\op{F}_{k}^{l}$ on all systems. Although both schemes are identical, viewing the feedback in this way simplifies our calculations.\label{fig:Derivation:MultiSystemFeedback}}
\end{figure}

Instead of a single system we now consider an array of systems $\sys{S}_{k}$, $k=1,\dots,N$, each coupled to a corresponding light field $\sys{A}_{k}$ as in Fig.~\hyperref[fig:Derivation:MultiSystemFeedback]{\ref{fig:Derivation:MultiSystemFeedback}~(a)}. Note that the systems need not be identical, and that there is no direct interaction between them.

As in the previous section we assume that the relevant system and light Hamiltonians allow us to change to an interaction frame in which the total system-light coupling can be expressed as
\begin{align}
	\op{H}_\mrm{int}(t) & = \sum_{j=1}^{N}\op{H}_\mrm{int}^{j}(t).
\end{align}
The Hamiltonians $\op{H}_\mrm{int}^{j}(t)$ generate the coupling of $\sys{S}_{j}$ to $\sys{A}_{j}$ and are given by
\begin{align}\label{eq:Derivation:InteractionHamiltonianManySystems}
	\op{H}_{\mrm{int}}^{j}(t) & = i\sqrt{\kappa}\left( \op{S}_{j}\hat{a}_{j}^{\dag}(t) - \op{S}_{j}^{\dag}\hat{a}_{j}(t)\right)
\end{align}
for every $j$. We assume here a common coupling strength $\kappa$ for all of the $N$ systems. Inhomogeneities of the coupling constants can be absorbed into the the system operators $\op{S}_{j}$.  Each $\op{S}_{j}$ is a local operator, meaning that it acts non-trivially only on system $\sys{S}_{j}$, and $\hat{a}_{j}(t)$ are bosonic annihilation operators as in Eq.~\eqref{eq:Derivation:InteractionHamiltonian}, obeying the usual relation
\begin{align}
	[\hat{a}_{j}(t), \hat{a}_{k}^{\dag}(t')] = \delta_{jk}\delta(t-t').
\end{align}

\subsubsection{Coarse-graining of time}
As in Section \ref{sec:Derivation:CoarseGrainingOfTime}, we coarse-grain the evolution by introducing the discrete time-step $\delta t$. We can adopt the corresponding operators and Hamiltonians from Eq.~\eqref{eq:Derivation:CoarseGraining},
\begin{subequations}
	\begin{align}
		\op{\bar{A}}_{k} & := \frac{1}{\sqrt{\delta t}}\int_{t}^{t+\delta t}\diff t'\, \hat{a}_{k}(t'), \\
		\op{\bar{H}}_{\mrm{int}} & = i\sum_{k=1}^{N} \left( \op{S}_{k}\op{\bar{A}}_{k}^\dag - \op{S}_{k}^\dag\op{\bar{A}}_{k}\right).
	\end{align}
\end{subequations}
For ease of notation we combine the operators $\op{\bar{A}}_{k}$ and $\op{S}_{k}$ into $N$-dimensional vectors $\vect{\bar{A}}$ and $\vect{S}$ respectively, so $\op{\bar{H}}_{\mrm{int}} = i( \vect{\op{S}}\cdot\vect{\op{\bar{A}}}^\dag - \vect{\op{S}}^\dag\cdot\vect{\op{\bar{A}}})$, with the dot-product $\vect{S}\cdot\vect{\bar{A}}^\dag := \sum\nolimits_{k}\op{S}_{k}\op{\bar{A}}_{k}^{\dag}$.

Initially, all incoming light fields are in the vacuum state, $\rho^{\sys{S}\sys{A}}(t) = \rho^{\sys{S}}(t)\otimes |\mrm{vac}\rangle \langle \mrm{vac}|$. Thus, as in Eq.~\eqref{eq:Derivation:InteractionDensMat}, the combined system-light state after the interaction will be
\begin{align}
	\rho_\mrm{int}^{\sys{S}\sys{A}} \approx \exp\left( -i\epsilon\op{\bar{H}}_\mrm{int} \right) \rho^{\sys{S}\sys{A}}(t) \exp\left( i\epsilon\op{\bar{H}}_\mrm{int} \right),
\end{align}
up to $\mcl{O}(\epsilon^2)$ where $\epsilon=\kappa\delta t$, as before.

\subsubsection{Interferometer and measurement}
The light beams traverse an interferometer after the interaction, which mixes the fields via a unitary $N\times N$ matrix $\mat{U}$, depicted by the blue box in Fig.~\hyperref[fig:Derivation:MultiSystemFeedback]{\ref{fig:Derivation:MultiSystemFeedback}~(a)}. The interferometer yields outgoing fields $\sys{B}_{1},\dots,\sys{B}_{N}$ with corresponding annihilation operators $\vect{\bar{B}} := \mat{U}\vect{\bar{A}}$. Changing the system operators in the same manner, so that
\begin{align}
	\vect{Z} & :=\mat{U}\vect{S},
\end{align}
and $\op{Z}_{k} := \sum\nolimits_{j}U_{kj}\op{S}_{j}$, allows us to rewrite the interaction,
\begin{align}
	\op{\bar{H}}_\mrm{int} = i(\vect{Z}\cdot\vect{\bar{B}}^\dag - \vect{Z}^\dag\cdot\vect{\bar{B}}).
\end{align}

We now perform a homodyne measurement of each $\sys{B}_{k}$. As before, we introduce the coarse-grained quadrature operators $\op{\bar{X}}_{k} := ( \op{\bar{B}}_{k} + \op{\bar{B}}_{k}^\dag )/\sqrt{2}$, and $\bar{P}_{k} := -i( \op{\bar{B}}_{k} - \op{\bar{B}}_{k}^\dag )/\sqrt{2}$ for $k=1,\dots,N$. Without loss of generality it suffices to measure only the $\op{\bar{X}}$-quadrature in all beams. Any other quadrature can be obtained by inserting phase plates before the measurement, which amounts to multiplying $\mat{U}$ with a diagonal phase matrix from the left. Since all $\op{\bar{X}}$-quadrature operators commute, we may define common eigenstates $|\vect{\bar{x}}\rangle:=|\bar{x}_{1},\dots,\bar{x}_{N}\rangle$ as usual through
\begin{align}
	\op{\bar{X}}_{k}|\vect{\bar{x}}\rangle = \bar{x}_{k}|\vect{\bar{x}}\rangle.
\end{align}
The conditional system state after the measurement carries over from Eq.~\eqref{eq:Derivation:ConditionalState},
\begin{align}
	\rho_\mrm{cond}^{\sys{S}}(\vect{\bar{x}}) := \frac{1}{p(\vect{\bar{x}})}\langle \vect{\bar{x}}| \rho_\mrm{int}^{\sys{S}\sys{A}}|\vect{\bar{x}} \rangle,
\end{align}
with normalization given by the joint p.d.f.~$p(\vect{\bar{x}})=\trace[S]{\langle \vect{\bar{x}}|\rho_\mrm{int}^{\sys{S}\sys{A}} |\vect{\bar{x}} \rangle}=\trace{|\vect{\bar{x}}\rangle\langle \vect{\bar{x}}| \rho_\mrm{int}^{\sys{S}\sys{A}}}$.

\subsubsection{Applying feedback}
As in Eq.~\eqref{eq:Derivation:FeedbackState}, we apply Hamiltonian feedback to the conditional state,
\begin{align}
	\rho_\mrm{fb}^{\sys{S}}(\vect{\bar{x}}) & = \expo{-i\sqrt{2}\epsilon\op{G}(\vect{\bar{x}})}\rho_\mrm{cond}^{\sys{S}}(\vect{\bar{x}})\expo{i\sqrt{2}\epsilon\op{G}(\vect{\bar{x}})},
\end{align}
with some Hermitian generator $\op{G}(\vect{\bar{x}})$ acting on the $N$ systems and depending on the measurement outcomes $\vect{\bar{x}}$. It is crucial to specify at this point the resources which are assumed in the feedback operations: in order to maintain the assumption that there is no direct physical interaction between the systems we restrict ourselves to local feedback operations, each affecting only individual systems as depicted in Fig.~\hyperref[fig:Derivation:MultiSystemFeedback]{\ref{fig:Derivation:MultiSystemFeedback}~(a)}. This means we can write $\op{G}(\vect{\bar{x}})=\sum\nolimits_{l=1}^{N} \op{G}_{l}(\vect{\bar{x}})$ with local operators $\op{G}_{l}(\vect{\bar{x}})$ acting (nontrivially) on system $l$ only. The feedback on system $\sys{S}_{l}$ will then be generated by $\op{G}_{l}(\vect{\bar{x}})=\sum\nolimits_{k=1}^{N}\bar{x}_{k}\op{F}_{k}^{l}$, where $\op{F}_{k}^{l}$ are some local Hermitian operators describing the effect of the measurement in channel $k$ on system $l$. Consequently, the total feedback on all systems yields
\begin{align}
	\op{G}(\vect{\bar{x}})=\sum\nolimits_{k,l=1}^{N}\bar{x}_{k}\op{F}_{k}^{l}. \label{eq:Derivation:SimultaneousFeedback}
\end{align}
Our choice of $\op{G}$ effects simultaneous action of all feedback operations. This is justified because, as we will show below Eq.~\eqref{eq:Derivation:MultiSysTrace}, different feedback operations commute up to $\mcl{O}(\epsilon^2)$ due to the independent Gaussian statistics of the light fields.

To simplify the notation, let us collect the feedback generated by each measurement $\bar{x}_{k}$,
\begin{align}
	\op{F}_{k} & := \sum_{l=1}^{N} \op{F}_{k}^{l},
\end{align}
as shown in Fig.~\hyperref[fig:Derivation:MultiSystemFeedback]{\ref{fig:Derivation:MultiSystemFeedback}~(b)}. If we combine all $\op{F}_{k}$ into a vector $\vect{\op{F}}$, we can write $\op{G}(\vect{\bar{x}})=\vect{\bar{x}}\cdot\vect{\op{F}}$ to obtain the feedback state
\begin{align}
	\rho_\mrm{fb}^{\sys{S}}(\vect{\bar{x}}) & = \expo{-i\sqrt{2}\epsilon\vect{\bar{x}}\cdot \vect{F}}\rho_\mrm{cond}^{\sys{S}}(\vect{\bar{x}})\expo{i\sqrt{2}\epsilon\vect{\bar{x}}\cdot \vect{F}}. \label{eq:Derivation:FeedbackStateMultiSys}
\end{align}
Note that Figs.~\hyperref[fig:Derivation:MultiSystemFeedback]{\ref{fig:Derivation:MultiSystemFeedback}~(a)} and \hyperref[fig:Derivation:MultiSystemFeedback]{\ref{fig:Derivation:MultiSystemFeedback}~(b)} describe the same setup, with only a slight change in notation and a different view on the feedback.

As in Eq.~\eqref{eq:Derivation:ConditionalFeedbackState}, we express $\rho_\mrm{fb}^{\sys{S}}(\vect{\bar{x}})$ in terms of the initial states of system and light,
\begin{align}\label{eq:Derivation:MultiSysFeedbackStateRewritten}
	\rho_\mrm{fb}^{\sys{S}}(\vect{\bar{x}}) & = \frac{1}{p(\vect{\bar{x}})} \langle \vect{\bar{x}}| \expo{-i\sqrt{2}\epsilon\vect{\op{\bar{X}}}\cdot \vect{\op{F}}} \rho_\mrm{int}^{\sys{S}\sys{A}} \expo{i\sqrt{2}\epsilon\vect{\op{\bar{X}}}\cdot \vect{\op{F}}} |\vect{\bar{x}} \rangle \notag\\
	& = \frac{1}{p(\vect{\bar{x}})} \langle \vect{\bar{x}}| \op{K}\left(\rho^\sys{S}\otimes \rho^{\sys{A}}\right) \op{K}^\dagger |\vect{\bar{x}} \rangle,
\end{align}
replacing $\bar{x}_{k}$ with $\op{\bar{X}}_{k}$, and setting
\begin{align}
	\op{K}:= \exp({-i\sqrt{2}\epsilon\vect{\op{\bar{X}}}\cdot\vect{F}}) \exp({-i\epsilon\op{\bar{H}}_\mrm{int}}).
\end{align}

\subsubsection{Unconditional master equation}
To obtain the unconditional system evolution we average the conditional state from Eq.~\eqref{eq:Derivation:MultiSysFeedbackStateRewritten} with respect to all light fields. This yields the familiar expression
\begin{align}
	\rho^{\sys{S}}(t+\delta t) & = \int_{\mathbb{R}^{N}} \diff \bar{x}_{1}\dots \diff \bar{x}_{N}\, p(\vect{\bar{x}})\rho_\mrm{fb}^{\sys{S}}(\vect{\bar{x}}) \notag\\
	& =  \trace[\sys{A}]{\op{K}\left(\rho^\sys{S}(t)\otimes |\mrm{vac}\rangle\langle\mrm{vac}|\right) \op{K}^\dagger}, \label{eq:Derivation:MultiSysTrace}
\end{align}
where $\trace[\sys{A}]{\dots}$ now denotes a partial trace over all light fields. We expand $K$ up to $\mcl{O}(\epsilon^2)$ to evaluate the trace. Dropping the superscript of the collective system state $\rho^{\sys{S}}\equiv\rho$, and using the independent Gaussian statistics of the light fields, $\langle \bar{x}_j \rangle = 0$ and $\langle \bar{x}_j \bar{x}_k \rangle = \delta_{jk}/2$, yields the density matrix
\begin{align}
	\rho(t+\delta t) & = \rho(t) -\frac{i}{2}\sum_{k=1}^{N}[\op{F}_{k}\op{Z}_{k} + \op{Z}_{k}^\dag \op{F}_{k},\rho(t)]\kappa\delta t \notag\\
	& \qquad + \sum_{k=1}^{N}\mcl{D}[\op{Z}_{k}-i\op{F}_k]\rho(t) \kappa\delta t.\label{eq:Derivation:MultiSysRho}
\end{align}

At this point we can justify the choice of simultaneous feedback in Eq.~\eqref{eq:Derivation:FeedbackStateMultiSys}. The order in which the feedback is applied is irrelevant for our purposes, as can be seen as follows: consider two generic feedback terms, $\sqrt{\delta t}\bar{x}_{k}\op{F}_{k}$ and $\sqrt{\delta t}\bar{x}_{k'}\op{F}_{k'}$. Using the Baker-Campbell-Hausdorff formula one finds
$\exp(\sqrt{\delta t}\bar{x}_{k}\op{F}_{k}) \exp(\sqrt{\delta t}\bar{x}_{k'}\op{F}_{k'}) = \exp(\sqrt{\delta t}\bar{x}_{k}\op{F}_{k} + \sqrt{\delta t}\bar{x}_{k'}\op{F}_{k'} + \frac{\delta t}{2}\bar{x}_{k}\bar{x}_{k'}[\op{F}_{k},\op{F}_{k'}]) + \mcl{O}(\delta t^{3/2})$. When performing the average from Eq.~\eqref{eq:Derivation:MultiSysTrace} the additional commutators $\bar{x}_{k}\bar{x}_{k'}[\op{F}_{k},\op{F}_{k'}]$ contribute only expectation values of the form $\langle \bar{x}_{k}\bar{x}_{k'} \rangle$. These vanish due to the uncorrelated Gaussian statistics of the vacuum fields, so our choice of feedback in Eq.~\eqref{eq:Derivation:SimultaneousFeedback} is justified.

As before, setting $\kappa=1$ in Eq.~\eqref{eq:Derivation:MultiSysRho} and taking the limit $\delta t \to \diff t$, we can read off the Feedback Master Equation (FME)
\begin{align}
	\dot{\rho} & = -\frac{i}{2}\sum_{k=1}^{N}[\op{F}_{k}\op{Z}_{k}+\op{Z}_{k}^\dag \op{F}_{k},\rho] + \sum_{k=1}^{N}\mcl{D}[\op{Z}_{k}-i\op{F}_{k}]\rho \notag\\
	& = -i[\op{H},\rho] + \sum_{k=1}^{N}\mcl{D}[\op{J}_{k}]\rho, \label{eq:Derivation:MultiSysMasterEquation}
\end{align}
with an effective Hamiltonian and jump operators
\begin{align}
		\op{H} & := \frac{1}{2}\sum_{k=1}^{N}(\op{F}_{k}\op{Z}_{k}+\op{Z}_{k}^\dag \op{F}_{k}),\tag{\ref*{eq:Derivation:MultiSysMasterEquation}a}\\
		J_{k} & := \op{Z}_{k}-i\op{F}_{k}, \tag{\ref*{eq:Derivation:MultiSysMasterEquation}b}
\end{align}
each comprising sums of local operators
\begin{align}
	\op{Z}_{k} = \sum_{j=1}^{N}U_{kj}\op{S}_{j}, & \qquad\text{and}\qquad \op{F}_{k} = \sum_{l=1}^{N}\op{F}_{k}^{l}. \tag{\ref*{eq:Derivation:MultiSysMasterEquation}c}
\end{align}
The FME for $N$ systems under continuous, diffusive, interferometric measurement and local feedback, Eq.~\eqref{eq:Derivation:MultiSysMasterEquation}, is the main result of this section. The resulting open-system many-body dynamics is determined by a Hamiltonian $\op{H}$ and a set of jump operators $\op{J}_k$, cf. Eqs.~(\hyperref[eq:Derivation:MultiSysMasterEquation]{\ref{eq:Derivation:MultiSysMasterEquation}a}) and (\hyperref[eq:Derivation:MultiSysMasterEquation]{\ref{eq:Derivation:MultiSysMasterEquation}b}), respectively. The Hamiltonian exhibits pairwise interactions of, in general, arbitrary range. The jump operators consist of sums of strictly local terms and, therefore, act collectively on the $N$ systems. Thus, the FME describes a fairly general class of open and interacting many-body dynamics of systems without requiring any direct physical interaction among them: all interactions are mediated by the interferometric measurement and feedback. These can be engineered almost arbitrarily in range or geometry by a proper choice of \textit{(i)} the system-light interactions characterized by the system operators $\op{S}_j$, cf.\ Eq.~\eqref{eq:Derivation:InteractionHamiltonianManySystems}, \textit{(ii)} the interferometer $\mat{U}$, and \textit{(iii)} the feedback scheme determined by the operators $\op{F}_k^l$, cf.\ Eq.~\eqref{eq:Derivation:SimultaneousFeedback}.  In the following sections we will consider possible choices for $\{\op{S}_j,\mat{U},\op{F}_k^l\}$, identify conditions for achieving non-trivial quantum dynamics, and provide particular models thereof.

\subsubsection{Generalization to multiple light modes\label{sec:Derivation:MultipleLightModes}}
As in the previous section, consider $N$ systems $\sys{S}_{k}$ coupled to fields $\sys{A}_{k}$, and let us denote the feedback dynamics from Eq.~\eqref{eq:Derivation:MultiSysMasterEquation} by
\begin{align}
	\dot{\rho} =  \mcl{L}^{\sys{A}}\rho := -i[\op{H}^{\sys{A}},\rho] + \sum_{k=1}^{N}\mcl{D}[\op{J}_{k}^{\sys{A}}]\rho.
\end{align}
Now assume that each $\sys{S}_{k}$ can simultaneously interact with a second field $\sys{B}_{k}$. These may correspond to different frequencies, polarizations, or spatial modes. We can construct a second interferometer $\mat{U}^{\sys{B}}$ for the $\sys{B}$-fields, and perform additional measurements and feedback to drive the systems. Turning off the $\sys{A}$-fields and considering only the dynamics generated by the $\sys{B}$-fields will generate an analogous feedback master equation
\begin{align}
	\dot{\rho} =  \mcl{L}^{\sys{B}}\rho := -i[H^{\sys{B}},\rho] + \sum_{k=1}^{N}\mcl{D}[J_{k}^{\sys{B}}]\rho.
\end{align}

If we take into account both, $\sys{A}$- and $\sys{B}$-fields, we find that the Liouvillians simply add up,
\begin{align}
	\dot{\rho} & = \mcl{L}^{\sys{A}}\rho + \mcl{L}^{\sys{B}}\rho.
\end{align}
The reason for this is the additive structure of the Hamiltonian and jump operators in Eqs.~\eqref{eq:Derivation:MultiSysMasterEquation}. Combine the system and feedback operators each into a single vector, $\vect{\op{S}} = (\vect{\op{S}}^\sys{A},\vect{\op{S}}^\sys{B})^\transpose$ and $\vect{\op{F}} = (\vect{\op{F}}^\sys{A},\vect{\op{F}}^\sys{B})^\transpose$. The new system operators after the interferometer are given by $\vect{\op{Z}} = (\vect{\op{Z}}^\sys{A},\vect{\op{Z}}^\sys{B})^\transpose :=  \mat{U}\vect{\op{S}}$, with a block matrix
\begin{align}
	\mat{U} =
	\begin{bmatrix}
		\mat{U}^\sys{A} & \mat{0} \\
		\mat{0} & \mat{U}^\sys{B}
	\end{bmatrix}.
\end{align}
Inserting this into Eqs.~\eqref{eq:Derivation:MultiSysMasterEquation} will yield a Hamiltonian $\op{H}=\op{H}^\sys{A} + \op{H}^\sys{B}$, and two sets of jump operators $\op{J}_{k}^{\sys{A}}$, $\op{J}_{k}^{\sys{B}}$, which results in the additive Liouvillians above. We can repeat this line of reasoning if we let the systems interact with additional fields.

\section{Parametrization of LOCC dynamics\label{sec:LOCC}}
Without an interferometer we would measure each mode separately and then perform local feedback on the systems through classical channels. Schemes of this type are known to generate only \textit{local operations and classical communication (LOCC)} dynamics \cite{Nielsen1999,Nielsen2010}, which may transform, but not produce any quantum correlations. Thus an interferometer is required if we wish to create entanglement. However, as we will show in this section, not any setup will do as one might end up in the LOCC regime even for a non-trivial interferometer.

\begin{figure}
	\includegraphics[width=\columnwidth,keepaspectratio=true]{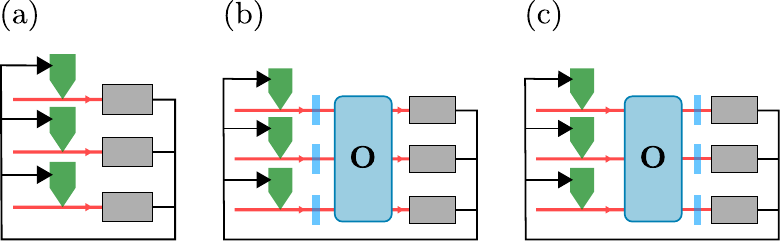}%
	\caption{(a) Without an interferometer we can perform only local measurements. In combination with local feedback, the scheme will generate only LOCC dynamics. (b) Local phase plates before an LOCC setup, such as a real orthogonal interferometer $\mat{O}$, will not enable any non-local operations. This is an LOCC setup which does not necessarily satisfy Eq.~\eqref{eq:LOCC:Condition}. (c) Phase plates before the detectors, on the other hand, change the homodyne measurement basis. They may enable non-LOCC dynamics, even if the preceding setup is LOCC. \label{fig:LOCC:NoIFO}}
\end{figure}

A setup without interferometer, as in Fig.~\hyperref[fig:LOCC:NoIFO]{\ref{fig:LOCC:NoIFO}~(a)}, corresponds to setting $\mat{U}$ equal to the $N\times N$ identity matrix, $\mat{U}=\mat{1}_N$. The resulting Feedback Master Equation (FME) from Eq.~\eqref{eq:Derivation:MultiSysMasterEquation} reads
\begin{equation}
	\dot{\rho} = -\frac{i}{2}\sum_{k=1}^{N}[ \op{F}_{k}\op{S}_{k} +\op{S}_{k}^\dag \op{F}_{k},\rho] +\sum_{k=1}^{N}\mcl{D}[\op{S}_{k}-i\op{F}_{k}]\rho, \label{eq:LOCC:LOCCFME}
\end{equation}
with local operators $\op{S}_{k}$, and feedback operators $\op{F}_{k}=\sum\nolimits_{l=1}^{N}\op{F}_{k}^{l}$. Although the FME retains its apparently non-local form, we know from the underlying setup that it will generate only LOCC dynamics, as will any master equation which can be written in this way.

Now consider the case of a non-trivial interferometer, $\mat{U}\neq\mat{1}_N$. We can always rewrite the general FME,
\begin{align}
	\dot{\rho} & = -\frac{i}{2}\sum_{k=1}^{N}[\op{F}_{k}\op{Z}_{k}+\op{Z}_{k}^\dag\op{F}_{k},\rho] + \sum_{k=1}^{N}\mcl{D}[\op{Z}_{k}-i\op{F}_{k}]\rho \notag\\
	& = -\frac{i}{2}\sum_{j=1}^{N}\left[ \op{\tilde{F}}_{j}\op{S}_{j} +\op{S}_{j}^\dag \op{\tilde{F}}_{j}^\dag,\rho\right] +\sum_{j=1}^{N}\mcl{D}[\op{S}_{j}-i\op{\tilde{F}}_{j}^\dag]\rho, \label{eq:LOCC:NonHermitianFeedbackFME}
\end{align}
by introducing formal (generally non-Hermitian) feedback operators
\begin{align}
	\op{\tilde{F}}_{j} & = \sum_{k=1}^{N}U_{kj} \op{F}_{k},
\end{align}
and using the invariance of Lindblad operators under unitary transformations,
\begin{align}
	\sum_{k}\mcl{D}[\op{J}_{k}]\rho & = \sum_{k}\mcl{D}\Big[\sum\nolimits_{l}U_{kl}\op{J}_{l}\Big]\rho.
\end{align}

If the $\op{\tilde{F}}_{j}$ all happen to be Hermitian, the rewritten master equation is equivalent to Eq.~\eqref{eq:LOCC:LOCCFME}, and thus also gives rise to LOCC dynamics. Hermiticity of $\op{\tilde{F}}_j$ can be formulated as
\begin{align}
	0 & = \op{\tilde{F}}_j - \op{\tilde{F}}_j^\dag \notag\\
	& = \sum_{k=1}^{N} (U_{kj}-U_{kj}^*) \op{F}_{k} \notag\\
	& = \sum_{l=1}^{N} \left[ \sum_{k=1}^{N} (U_{kj}-U_{kj}^*)\op{F}_{k}^{l} \right].
\end{align}
Since each term in the sum over $l$ acts on a different system, they are linearly independent and must vanish individually. Thus a sufficient condition for LOCC dynamics reads
\begin{equation}
	\sum_{k=1}^{N} \mrm{Im}[U_{kj}]\op{F}_{k}^{l} = 0,\quad \text{for all}\ j,l = 1,\dots,N. \label{eq:LOCC:Condition}
\end{equation}
In particular, this condition is satisfied for any real orthogonal interferometer $\mat{U} = \mat{O}$, which mixes beams without introducing a relative phase shift. The reason is that a general complex $\mat{U}$ mixes $\hat{x}$- and $\hat{p}$-quadratures, which creates non-commuting observables, while a real orthogonal $\mat{U}=\mat{O}$ does not.

However, Eq.~\eqref{eq:LOCC:Condition} is not a necessary condition for LOCC dynamics. For example, consider the diagonal matrix $\mat{U}=\mat{\Lambda}:=\diag{\expo{i\theta_1},\dots,\expo{i\theta_{N}}}$, which causes local phase shifts without mixing the fields, or any real orthogonal matrix $\mat{O}$ with phases multiplied from the right, $\mat{U}=\mat{O}\mat{\Lambda}$, see Fig.~\hyperref[fig:LOCC:NoIFO]{\ref{fig:LOCC:NoIFO}~(b)}. In both configurations, absolute phases are imprinted on the incoming fields which can be absorbed into the system operators $\op{S}_{k}$. They constitute LOCC setups that need not satisfy Eq.~\eqref{eq:LOCC:Condition}.

On the other hand, the seemingly similar case of multiplication of $\mat{O}$ with phases from the left, $\mat{U}=\mat{\Lambda} \mat{O}$, actually changes the homodyne measurement basis, see Fig.~\hyperref[fig:LOCC:NoIFO]{\ref{fig:LOCC:NoIFO}~(c)}. This may facilitate non-LOCC dynamics, even if the interferometer alone does not. This fact is made use of in Section \ref{sec:StatePreparation:2QP}.

\section{Dissipative state preparation}
\label{sec:StatePreparation}
In this section we examine the feedback scheme with regard to dissipative state engineering \cite{Kraus2008}, where the goal is to generate nontrivial quantum correlations in the stationary state achieved in the long-time limit of the FME in Eq.~\eqref{eq:Derivation:MultiSysMasterEquation}. One particular example of this kind has been given recently by Hofer \textit{et al.}\ \cite{Hofer2013} who showed that it is possible to design a feedback master equation which deterministically drives two two-level-systems (qubits) into an entangled state. We begin in Section \ref{sec:StatePreparation:2QP} by reviewing this two-qubit protocol, which serves as a prime illustration of how the general formalism presented in the previous section might be employed. In Section \ref{sec:StatePreparation:Extension} we extend it to include more than two qubits, and show in Section \ref{sec:StatePreparation:Results} that it can produce entangled many-body states.

\subsection{Two-qubit protocol}
\label{sec:StatePreparation:2QP}
\begin{figure}
	\includegraphics[width=\columnwidth,keepaspectratio=true]{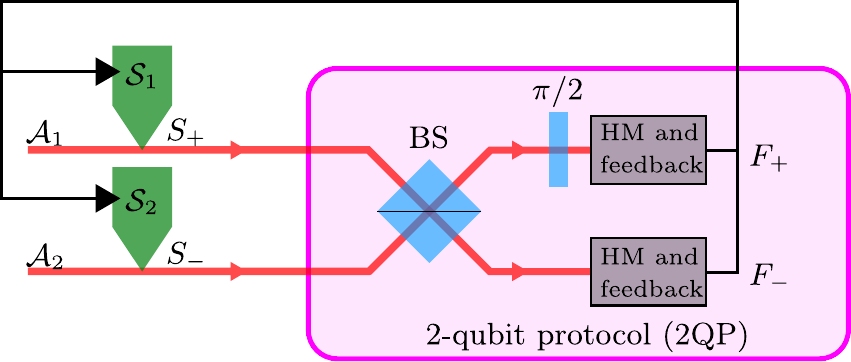}%
	\caption{Illustration of the two-qubit protocol (2QP). Both qubits $\sys{S}_j$ interact with light fields via $\op{S}_{+}$ or $\op{S}_{-}$ respectively. The fields are superposed on a balanced beam splitter (BS), after which one beam is phase-shifted by $\pi/2$. We perform a homodyne measurement (HM) of each field, and apply feedback $\op{F}_{\pm}$ proportional to the measured signal. The qubits will eventually relax into the steady state $|\psi(z)\rangle$. \label{fig:2QP:2QP}}
\end{figure}

The original two-qubit protocol is depicted in Fig.~\ref{fig:2QP:2QP}. It comprises two physical qubits, $\sys{S}_{1}$ and $\sys{S}_{2}$, two light fields, $\sys{A}_{1}$ and $\sys{A}_{2}$, and a balanced beam splitter mixing the beams. Subsequently, the phase of one beam is shifted by $\pi/2$. Together with the homodyne detection this realizes a continuous Bell measurement, where $\op{X}_{1}=\op{X}_{\sys{A}_{1}}+\op{X}_{\sys{A}_{2}}$ is measured in the top detector and the bottom detector measures $\op{X}_{2}=\op{P}_{\sys{A}_{1}} - \op{P}_{\sys{A}_{2}}$. Here, $\op{X}_{\sys{A}_{j}}$ ($\op{P}_{\sys{A}_{j}}$) denotes the amplitude (phase) quadrature of incoming field $\sys{A}_{j}$. Choosing appropriate feedback operators then drives the two qubits into the state $|\psi(z)\rangle \propto |00\rangle - z|11\rangle$. The parameter $z\in (0,1)$ is chosen beforehand to fix the system-light coupling and feedback gains, which then determine $|\psi(z)\rangle$.

The system-light coupling is generated by the system operators
\begin{equation}
	\begin{aligned}
		S_{+} & = s_{+}\hat{\sigma}_{+}^{1} + s_{-}\hat{\sigma}_{-}^{1},\\
		S_{-} & = s_{+}\hat{\sigma}_{+}^{2} - s_{-}\hat{\sigma}_{-}^{2},
	\end{aligned}
\end{equation}
with $s_{+}=\sqrt{z(1+z)}$ and $s_{-}=\sqrt{1-z}$ \footnote{How this specific system-light coupling might be achieved experimentally is outlined in Appendix D of \cite{Hofer2013}.}. Here, $\hat{\sigma}_{\pm}^{j}$ denote the standard Pauli raising and lowering operators acting on the $j$th qubit. We differ slightly from the notation of Section \ref{sec:Derivation:Generalization}, replacing the system operators $\op{S}_{1}$ and $\op{S}_{2}$ with $\op{S}_{+}$ and $\op{S}_{-}$ respectively, to more easily include multiple qubits in the following section.

The beam splitter and subsequent phase plate are represented by
\begin{align}
	\mat{U} & = \frac{1}{\sqrt{2}}\begin{bmatrix} 1 & 0 \\ 0 & i \end{bmatrix} \begin{bmatrix} 1 & 1 \\ 1 & -1	\end{bmatrix} = \frac{1}{\sqrt{2}}\begin{bmatrix} 1 & 1 \\ i & -i	\end{bmatrix},
\end{align}
so that the new system operators read
\begin{equation}
	\begin{aligned}
		Z_{+} & = \frac{1}{\sqrt{2}}(S_{+}+S_{-}), \\
		Z_{-} & = \frac{i}{\sqrt{2}}(S_{+}-S_{-}).
	\end{aligned}
\end{equation}
We choose the feedback operators
\begin{equation}
	\begin{aligned}
		F_{+} & = g_{+}\hat{\sigma}_{y}^{1} + g_{-}\hat{\sigma}_{y}^{2},\\
		F_{-} & = g_{-}\hat{\sigma}_{x}^{1} - g_{+}\hat{\sigma}_{x}^{2}, \label{eq:StatePreparation:2QP:SingleFeedback}
	\end{aligned}
\end{equation}
with gain coefficients $g_{\pm} = z(s_{-}\pm s_{+})/(\sqrt{2}s_{+}s_{-})$. The resulting system master equation, Eq. \eqref{eq:Derivation:MultiSysMasterEquation}, then reads
\begin{align}
	\dot{\rho} = \mcl{L}\rho & := -\frac{i}{2}[F_{+}Z_{+} + F_{-}Z_{-} + \mrm{h.c.}, \rho ] \notag\\
	&\qquad + \mcl{D}[\op{J}_{+}]\rho + \mcl{D}[\op{J}_{-}]\rho, \label{eq:StatePreparation:2QP:2QPFME}
\end{align}
with jump operators $\op{J}_{\pm}=\op{Z}_{\pm}-i\op{F}_{\pm}$.

The operators $\op{F}_{\pm}$ and $\op{Z}_{\pm}$ were chosen such that the master equation has the unique stationary state
\begin{align}
	|\psi(z) \rangle & = \frac{1}{\sqrt{1+z^2}}(|00\rangle - z|11\rangle), \label{eq:StatePreparation:2QP:SteadyState}
\end{align}
which can be seen as follows. The jump operators are
\begin{equation}
	\begin{aligned}
		\op{J}_{+} & \propto \op{K}_{1} + \lambda(z) \op{K}_{2},\\
		\op{J}_{-} & \propto \lambda(z)\op{K}_{1} - \op{K}_{2},
	\end{aligned}
\end{equation}
with $\op{K}_{1} = \hat{\sigma}_{-}^{1} + z\hat{\sigma}_{+}^{2}$ and $\op{K}_{2} = \hat{\sigma}_{-}^{2} + z\hat{\sigma}_{+}^{1}$, for some $\lambda(z)\in\mathbb{R}$. The operators $\op{K}_{1}$ and $\op{K}_{2}$ have $|\psi(z) \rangle$ as their common \textit{dark state} \cite{Kraus2008}, i.e., as eigenstate with eigenvalue 0. The particular combination of $\op{K}_{1}$ and $\op{K}_{2}$ that makes up $\op{J}_{+}$ and $\op{J}_{-}$ was chosen to realize $|\psi(z) \rangle$ also as eigenstate of the Hamiltonian. These two properties place $|\psi(z) \rangle\langle \psi(z)|$ in the kernel of Liouvillian $\mcl{L}$, and make it a steady state of the system \cite{Kraus2008}. Further investigation shows that $|\psi(z) \rangle$ is the \textit{only} steady state, which entails that regardless of the initial state, the qubits will eventually be driven into $|\psi(z)\rangle$. In other words, $|\psi(z)\rangle$ is prepared deterministically after some finite relaxation time.

In the following we will treat $z$ as a variable rather than as a constant to be fixed at the start. This is justified because when $z$ is changed, $z\to z'$ say, the qubits will relax into the new steady state $|\psi(z')\rangle$. If the variation of $z$ is sufficiently slow compared to the relaxation time, we expect the qubits to adiabatically follow the state space trajectory $\{|\psi(z)\rangle,\,z\in[0,1)\}$.

The interesting result of \cite{Hofer2013}, which also inspired the present work, was the deterministic preparation of entanglement. For any $0<z<1$ we see that $|\psi(z)\rangle$ is entangled, and will ideally approach the maximally entangled Bell state $|\Phi^{-}\rangle\propto|00\rangle - |11\rangle$ as $z\to 1$. This ideal limit can never be realized exactly since it leads to infinite feedback gains, $|g_{\pm}|\to\infty$. But using a more realistic description of the dynamics including passive photon loss \cite{Hofer2013,Hofer2015}, the highest possible degree of entanglement is always obtained for $z<1$. Keeping $s_{\pm}$ as above while optimizing $g_{\pm}$ to maximize the amount of entanglement in the steady state for every $z$, the optimal gains $g_{\pm}$ always remain finite, and one finds an entangled steady state for up to $50\%$ photon loss.

The opposite limit of $z=0$ corresponds to a setup without feedback, which entails $\op{H}=0$. With jump operators $\op{J}_{k} \sim \op{S}_{k} = \hat{\sigma}_{-}^{k}$, the qubits are incoherently pumped into the trivial state $|00\rangle$.

\subsection{Extension to multiple qubits\label{sec:StatePreparation:Extension}}
\begin{figure}
	\includegraphics[width=\columnwidth,keepaspectratio=true]{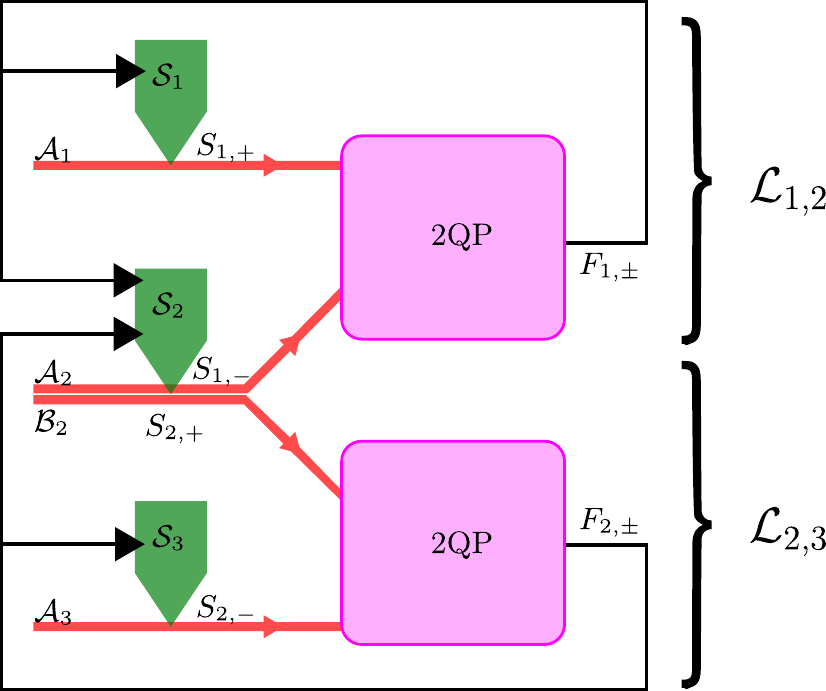}%
	\caption{The two-qubit protocol (2QP) is extended to 3 qubits. This is done by letting two light fields, $\sys{A}_{2}$ and $\sys{B}_{2}$, interact with $\sys{S}_2$, and using a second copy of the 2QP. $\sys{A}_{2}$ interacts via $\op{S}_{1,-}$ and then enters the first 2QP with light coming from $\sys{S}_1$. This generates feedback $\op{F}_{1,\pm}$ on $\sys{S}_1\sys{S}_2$ which gives rise to the Liouvillian $\mcl{L}_{1,2}$. Analogously, $\sys{B}_{2}$ interacts with $\sys{S}_2$ via $\op{S}_{2,+}$ and then enters the second 2QP. This results in feedback $\op{F}_{2,\pm}$ on $\sys{S}_2\sys{S}_3$ and Liouvillian $\mcl{L}_{2,3}$. \label{fig:2QP:MultiSystem}}
\end{figure}

Motivated by this deterministic preparation of a non-trivial quantum state of qubits, we examined whether and how the scheme could be extended to generate multipartite entangled states. To begin, consider a third qubit $\sys{S}_{3}$ added to the existing setup, and coupled to $\sys{S}_{2}$ via a replica of the original protocol with the same parameter $z$, shown in Fig.~\ref{fig:2QP:MultiSystem}. This is realized by letting $\sys{S}_{2}$ simultaneously interact with two light fields, $\sys{A}_{2}$ and $\sys{B}_{2}$, e.g., different polarizations of the same field or different spatial modes, as discussed in Section \ref{sec:Derivation:MultipleLightModes}.

We now generalize the operators $\op{S}_{\pm}$ and $\op{F}_{\pm}$ from the previous section. Let the index $j=1,2$ label the different two-qubit-setups, each coupling a pair $\sys{S}_{j}\sys{S}_{j+1}$. The system operators for the $j$th setup read
\begin{equation}
	\begin{aligned}
		\op{S}_{j,+} & = s_{+}\hat{\sigma}_{+}^{j} + s_{-}\hat{\sigma}_{-}^{j}, \\
		\op{S}_{j,-} & = s_{+}\hat{\sigma}_{+}^{j+1} - s_{-}\hat{\sigma}_{-}^{j+1},
	\end{aligned}
\end{equation}
and are mixed by the beam splitter and phase plate of each setup to yield
\begin{equation}
	\begin{aligned}
		\op{Z}_{j,+} & = (\op{S}_{j,+}+\op{S}_{j,-})/\sqrt{2}, \\
		\op{Z}_{j,-} & = i(\op{S}_{j,+}-\op{S}_{j,-})/\sqrt{2}.
	\end{aligned}
\end{equation}
The feedback is mediated via operators
\begin{equation}
	\begin{aligned}
		\op{F}_{j,+} & = g_{+}\hat{\sigma}_{y}^{j} + g_{-}\hat{\sigma}_{y}^{j+1}, \\
		\op{F}_{j,-} & = g_{-}\hat{\sigma}_{x}^{j} - g_{+}\hat{\sigma}_{x}^{j+1}. \label{eq:StatePreparation:2QP:MultiFeedback}
	\end{aligned}
\end{equation}

The resulting Liouvillian $\mcl{L}_{j,j+1}$ acts on the pair $\sys{S}_{j}\sys{S}_{j+1}$ as indicated in Fig.~\ref{fig:2QP:MultiSystem}. It takes the same form as in Eq.~\eqref{eq:StatePreparation:2QP:2QPFME},
\begin{align}
	\mcl{L}_{j,j+1}\rho & := -\frac{i}{2}[\op{F}_{j,+}\op{Z}_{j,+} + \op{F}_{j,-}\op{Z}_{j,-} + \mrm{h.c.}, \rho ] \notag\\
	&\qquad + \mcl{D}[\op{J}_{j,+}]\rho + \mcl{D}[\op{J}_{j,-}]\rho,
\end{align}
with the proper system and feedback operators, and jump operators $\op{J}_{j,\pm}=\op{Z}_{j,\pm}-i\op{F}_{j,\pm}$. The total feedback master equation for all three qubits then comprises both Liouvillians,
\begin{align}
	\dot{\rho} & = \mcl{L}_{1,2}\rho + \mcl{L}_{2,3}\rho.
\end{align}

It is now straightforward to extend the setup and formalism to $N$ qubits $\sys{S}_{1},\dots,\sys{S}_{N}$ with corresponding $\mcl{L}_{j,j+1}$ for $j=1,\dots,N-1$. This represents an open-ended chain with nearest-neighbor interaction. We can turn this chain into a ring by also coupling $\sys{S}_{N}$ to $\sys{S}_{1}$ via $\mcl{L}_{N,N+1}\equiv \mcl{L}_{N,1}$. This gives rise to the master equation
\begin{align}
	\dot{\rho} & = \sum_{j=1}^{N} \mcl{L}_{j,j+1}\rho,
\end{align}
which generates periodic and translationally invariant dynamics. The steady state should inherit these symmetries.

Recall that the non-periodic two-qubit version of the protocol prepares the state $|\psi(z) \rangle$ from Eq.~\eqref{eq:StatePreparation:2QP:SteadyState}, which goes from a product state for $z=0$ to a maximally entangled state as $z\to 1$. It was shown that bipartite entanglement in a system of qubits %, as measured by the concurrence \cite{Coffman2000},
is monogamous \cite{Coffman2000,Osborne2006}. Consequently, highly entangling two qubits, $\sys{S}_{1}$ and $\sys{S}_{2}$ say, places an upper bound on the possible entanglement between $\sys{S}_{2}$ and $\sys{S}_{3}$ (and so on). Hence we expect some competition between the pairwise dynamics generated by each $\mcl{L}_{j,j+1}$, reminiscent of a frustrated spin chain.

\subsection{Analysis of the steady state\label{sec:StatePreparation:Results}}
The competing Liouvillians $\mcl{L}_{j,j+1}$ give rise to a non-trivial collective steady state of the qubits. To investigate it we use the Python package ``QuTiP'' \cite{Johansson2012,Johansson2013a} to obtain exact results for small numbers of systems ($N\leq 7$), and a recently established variational procedure \cite{Weimer2015} in the limit of a large number of systems.

The central element of the variational procedure is a variational principle for the steady state of an open quantum system. After restricting the steady state ansatz to a certain variational manifold, the steady state equation $(\diff/\diff t) \rho = 0$ can no longer be solved, as the true steady state will generically lie outside the variational manifold. Therefore, the variational norm $||(\diff/\diff t) \rho||$ is minimized instead to obtain an approximation for the steady state. Importantly, the choice of the variational norm is not arbitrary, but has to be the trace norm $||(\diff/\diff t) \rho|| = \trace{|(\diff/\diff t) \rho|}$ \cite{Weimer2015,Weimer2015a}.

Here, we parametrize the density operator according to
\begin{align}
	\rho &= \rho_0 \otimes \rho_0 \otimes \cdots  + \sum\limits_{i} \rho_0 \otimes \cdots \otimes C_{i,i+1} \otimes \rho_0 \cdots\\ &+ \sum\limits_{i,j}\rho_0 \otimes \cdots \otimes C_{i,i+1}\otimes \rho_0 \cdots \otimes  C_{j,j+1} \otimes \rho_0\cdots+\ldots\nonumber
\end{align}
The first term is simply a product state of all qubits being in the state $\rho_0$. The other terms involve the nearest-neighbor correlation matrices $C_{i,i+1}$ and allow to describe nonclassical correlations such as entanglement. As the calculation of the exact variational norm is in general still an intractable problem, we resort to an upper bound to the norm that can be efficiently calculated, and is given by a sum of three-qubit problems, $||(\diff/\diff t) \rho||\leq \sum_i ||(\diff/\diff t) \rho_{i-1,i,i+1}||$, where $\rho_{i-1,i,i+1}$ is the reduced
density operator involving three qubits \cite{Weimer2015,Weimer2015a}. As we consider a translationally invariant problem, minimizing a single three-qubit term minimizes the full sum as well.

\subsubsection*{Results}
\begin{figure}
	\includegraphics[width=\columnwidth,keepaspectratio=true]{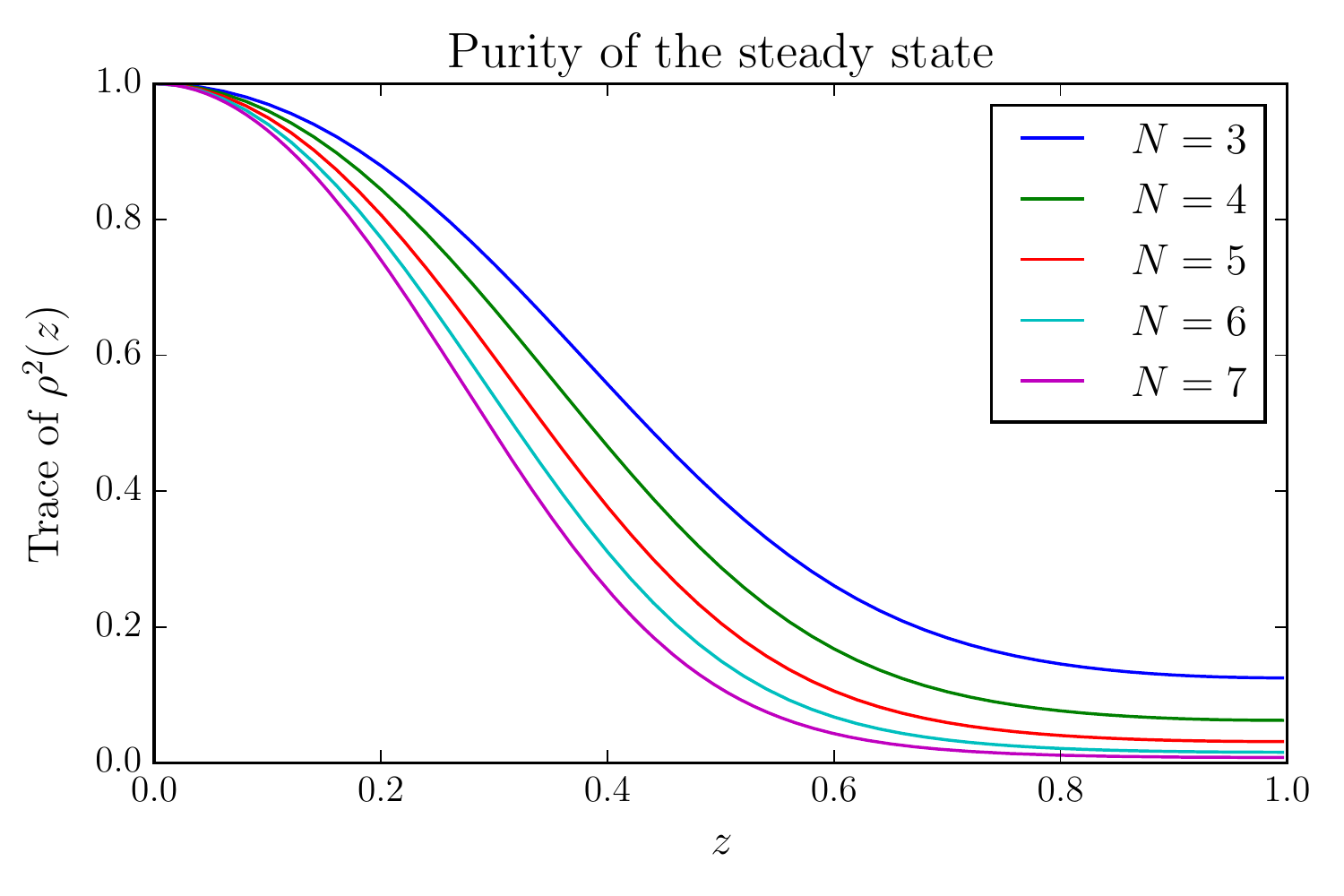}%
	\caption{The purity of the steady state varies with $z$. Without feedback ($z=0$) all systems are incoherently pumped into $|0\rangle$. As $z$ grows, however, competition between the pairwise dynamics increasingly mixes the steady state. The maximally mixed state is approached with $\trace{\rho^2(z)}\to 2^{-N}$ as $z\to 1$. The curves in the legend are shown from top to bottom in the figure.\label{fig:2QP:Purity}}
\end{figure}

As before, turning off the feedback ($z=0$) yields the trivial pure state
\begin{align}
	|\psi(z=0)\rangle = \bigotimes_{j=1}^{N} |0\rangle_{j},
\end{align}
obtained by incoherent pumping of the individual systems. With feedback ($z>0$), however, the competing dynamics immediately take effect. This can be seen in Fig.~\ref{fig:2QP:Purity}, where we plot the trace of $\rho^2(z)\equiv (\rho(z))^2$ to quantify the \textit{purity} of the steady state for different values of $z$. As $z$ grows, $\rho(z)$ becomes increasingly mixed and approaches the maximally mixed state for $z\to 1$, where the feedback gains become infinite. Thus we expect to find no quantum correlations in either of the limits $z\to 0$ or $z\to 1$.

\begin{figure}
	\includegraphics[width=\columnwidth,keepaspectratio=true]{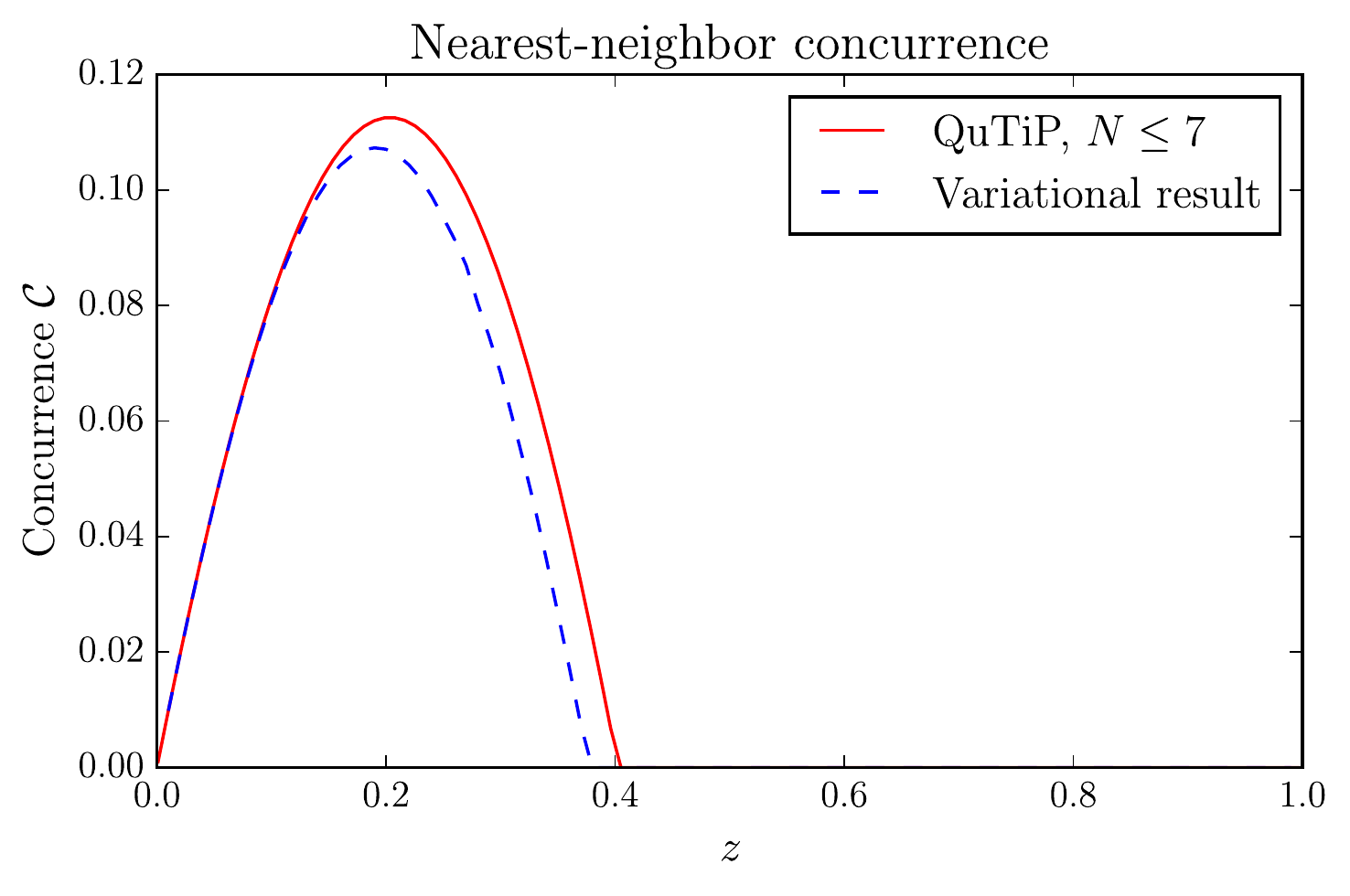}%
	\caption{The concurrence $\mcl{C}(\rho_{k,k+1}(z))$ of neighboring qubits in the steady state is non-zero for $0<z<0.4$, showing the presence of entanglement. We computed the steady state directly using the Python package QuTiP \cite{Johansson2012,Johansson2013a} for small system size ($N\leq 7$ qubits), and found the concurrence to be independent of $N$ (solid red line). For large $N$, we analyzed the steady state using a variational approach \cite{Weimer2015}. While the variational concurrence (dotted blue line) takes on slightly different values, it is still in good qualitative agreement with our exact results. We found the concurrence between non-neighboring qubits to vanish everywhere. \label{fig:2QP:Concurrence}}
\end{figure}

\begin{figure}
	\includegraphics[width=\columnwidth,keepaspectratio=true]{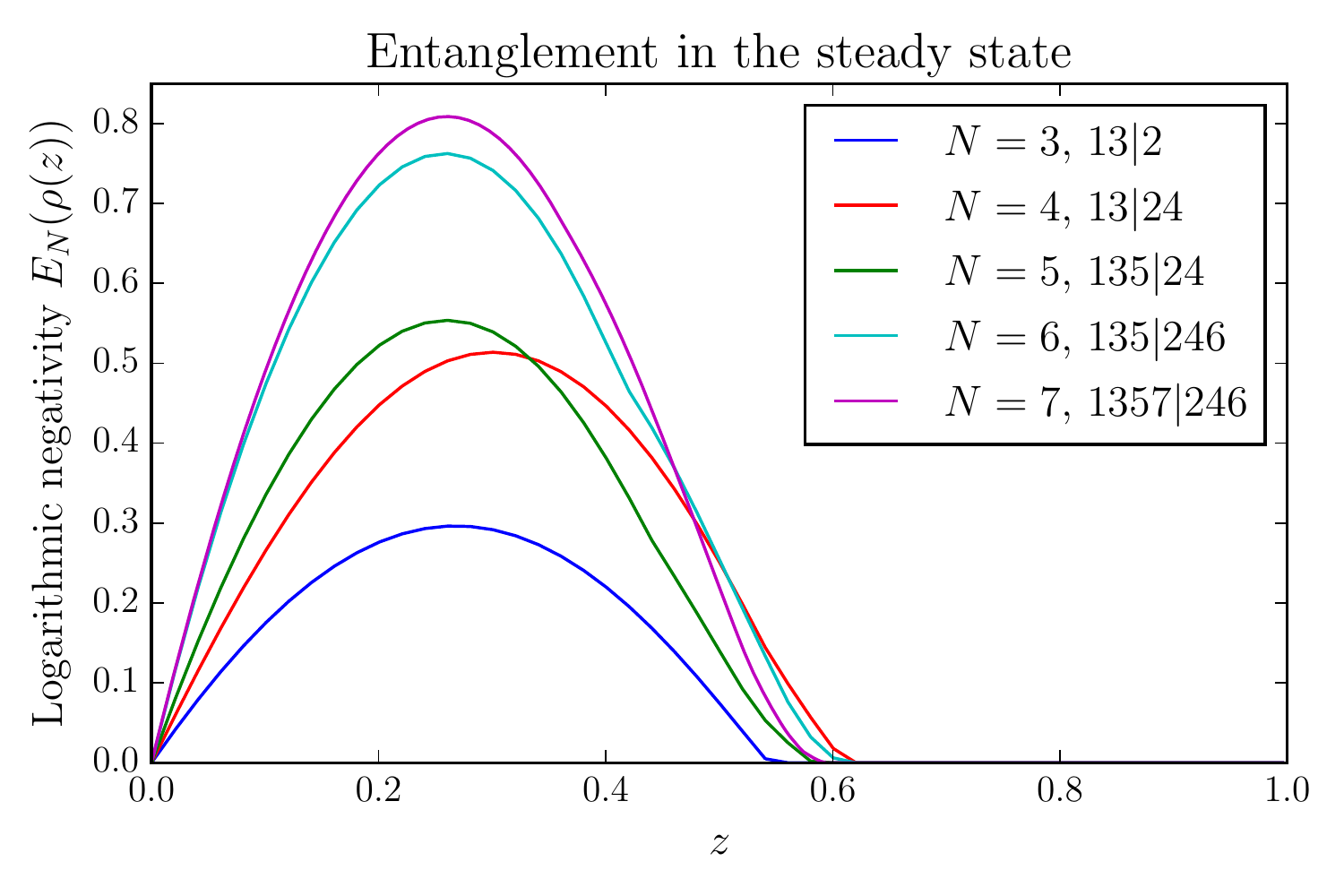}%
	\caption{The logarithmic negativity $E_{N}(\rho(z))$ in the steady state for ``odd$|$even''-bipartitions of setups comprising $N=3,\dots,7$ qubits. Increasing the system size causes a larger peak negativity. Notably, the entanglement persists for values of around $z\approx 0.6$, where the concurrence has long vanished. The curves in the legend are shown from bottom to top in the figure.\label{fig:2QP:LogNegativity}}
\end{figure}

In the regime $0<z<1$, on the other hand, entanglement does form, as can be seen in Figs.~\ref{fig:2QP:Concurrence} and \ref{fig:2QP:LogNegativity}. A simple indicator for bipartite entanglement is the \textit{concurrence} $\mcl{C}(\rho_{kl}(z))$ \cite{Coffman2000,Osborne2006,Plenio2006}, which is related to the entanglement of formation for two-qubit states. It can be computed for arbitrary pairs of qubits $\sys{S}_{k}\sys{S}_{l}$ after taking the partial trace over all other systems, $\rho_{kl}(z):=\trace[\neq \sys{S}_{k},\sys{S}_{l}]{\rho(z)}$. For the two-qubit state $\rho_{kl}$ the concurrence is then defined as
\begin{align}
	\mcl{C}(\rho_{kl}) = \max\{0,\lambda_{1}-\lambda_{2}-\lambda_{3}-\lambda_{4}\},
\end{align}
where the $\lambda_{j}$ are the square roots of the eigenvalues of $\rho_{kl} (\hat{\sigma}_{y}\otimes \hat{\sigma}_{y})\rho_{kl}^*(\hat{\sigma}_{y}\otimes \hat{\sigma}_{y})$ in decreasing order.

Since the dynamics, and hence the steady state, is translationally invariant and periodic, the concurrence only depends on the distance between two qubits on the ring. Our analysis revealed that the concurrence of nearest neighbors is non-zero in the region $0<z<0.4$, see Fig.~\ref{fig:2QP:Concurrence}. Thus the steady state is indeed entangled in this regime. We found that the concurrence is independent of the number of systems for $N\leq 7$, and qualitatively agrees with the variational results. On the other hand, the concurrence between non-neighboring qubits vanishes everywhere, which may be because each Liouvillian $\mcl{L}_{j,j+1}$ acts only on neighboring pairs.

Another quantity we considered (see Fig.~\ref{fig:2QP:LogNegativity}) is the \textit{logarithmic negativity} \cite{Vidal2002,Plenio2005a,Plenio2006},
\begin{align}
	E_{N}(\rho_{\sys{X}\sys{Y}}) = \log_{2}\|\rho_{\sys{X}\sys{Y}}^{\transpose_{\sys{X}}} \|_1,
\end{align}
where $\rho_{\sys{X}\sys{Y}}$ is the state of a bipartite system $\sys{X}|\sys{Y}$, the operation $\transpose_{\sys{X}}$ denotes the partial transpose with respect to subsystem $\sys{X}$, and $\|\rho\|_{1}$ is the sum of singular values of $\rho$. Neither of the subsystems, $\sys{X}$ or $\sys{Y}$, need to be qubits, and may each constitute multipartite systems themselves. If $E_{N}(\rho_{\sys{X}\sys{Y}})$ is non-zero, systems $\sys{X}$ and $\sys{Y}$ are entangled, so it provides a sufficient (but not necessary \cite{Vidal2002,Plenio2005a}) condition for the presence of quantum correlations. The largest region in which we were able to detect entanglement in this way was for up to $z\approx 0.6$. Out of all possible bipartitions, the entanglement always extended furthest for the ``odd$|$even''-partition with subsystems $\sys{X}=\sys{S}_{1}\sys{S}_{3}\dots$ and $\sys{Y}=\sys{S}_{2}\sys{S}_{4}\dots$.

It is interesting that the entanglement as measured by the logarithmic negativity not only extends further but also peaks later than the concurrence. This may be attributed to the fact that the concurrence is, in a sense, a more local quantity since it detects only correlations between two qubits, while the logarithmic negativity takes into account the collective $N$-qubit state. This difference in behavior might indicate that the quantum correlations are not simply destroyed as $z$ grows, but instead manifest in more complex or long-ranged form, such as true multipartite entanglement \cite{Vidal2002,Plenio2006}.

These results settle the question of whether the feedback scheme can reliably produce entangled many-body states. Both the concurrence and the logarithmic negativity provide clear evidence for quantum correlations in the steady state, which could not necessarily be expected from the underlying setup or the master equation alone. Furthermore, the entanglement appears to be present independent of the number of systems. 
\section{Quantum simulation\label{sec:QuantumSimulation}}
We now approach the feedback setup from the perspective of quantum simulation. It can be difficult to design and control many-body quantum systems in experiments because of their complex interaction, and it is often not feasible to simulate them on classical computers due to their large state space. A full-scale quantum computer  might relieve these issues, but is still out of reach. This gap is filled by quantum simulation \cite{Feynman1982,Lloyd1996,Buluta2009,Georgescu2014,Houck2012,Lanyon2011,Kim2010,Porras2004}, where a complex physical many-body system is emulated using a more easily controllable setup.

Our feedback scheme may prove useful in this regard since it realizes pairwise interaction and collective dissipation between distant systems, while using only standard techniques of quantum optics. To gauge its scope we realize a dissipative Ising spin model with local transverse fields. Such models are the subject of active research because of their relation to ultracold Rydberg atoms \cite{Hu2013,Ates2012,Lee2011,Marcuzzi2014,Hoening2014}.

\subsection{Open Ising model with transverse fields\label{sec:QuantumSimulation:Ising}}
The system we wish to emulate comprises $N$ interacting spin-1/2 particles with Hamiltonian
\begin{align}
	\op{H}_{\text{Ising}} & = \sum_{k\neq l}^{N} \Delta_{kl}\hat{\sigma}_{x}^{k} \hat{\sigma}_{x}^{l} - \sum_{k=1}^{N}B_{k} \hat{\sigma}_{z}^{k}, \label{eq:IsingModel:Hamiltonian}
\end{align}
where $\hat{\sigma}_{u}^{j}$ ($u=x,y,z$) denotes a Pauli operator acting on the $j$th system. The parameters $\Delta_{kl}\in\mathbb{R}$ determine the interaction between systems $k$ and $l$, while $B_{j}\in\mathbb{R}$ is the strength of local magnetic fields. If we set $\Delta_{kk}=-B_{k}$, all parameters can be encoded in a real matrix $\mat{\Delta}=(\Delta_{kl})$. Note that we do not make assumptions about the range or geometry of the interaction, so $\op{H}_{\mrm{Ising}}$ may apply to, e.g., a chain, a ring, or a $d$-dimensional lattice, depending solely on the choice of $\mat{\Delta}$.

\subsubsection{Engineering the Hamiltonian\label{sec:QuantumSimulation:Ising:Hamiltonian}}
\begin{figure}
	\includegraphics[width=\columnwidth,keepaspectratio=true]{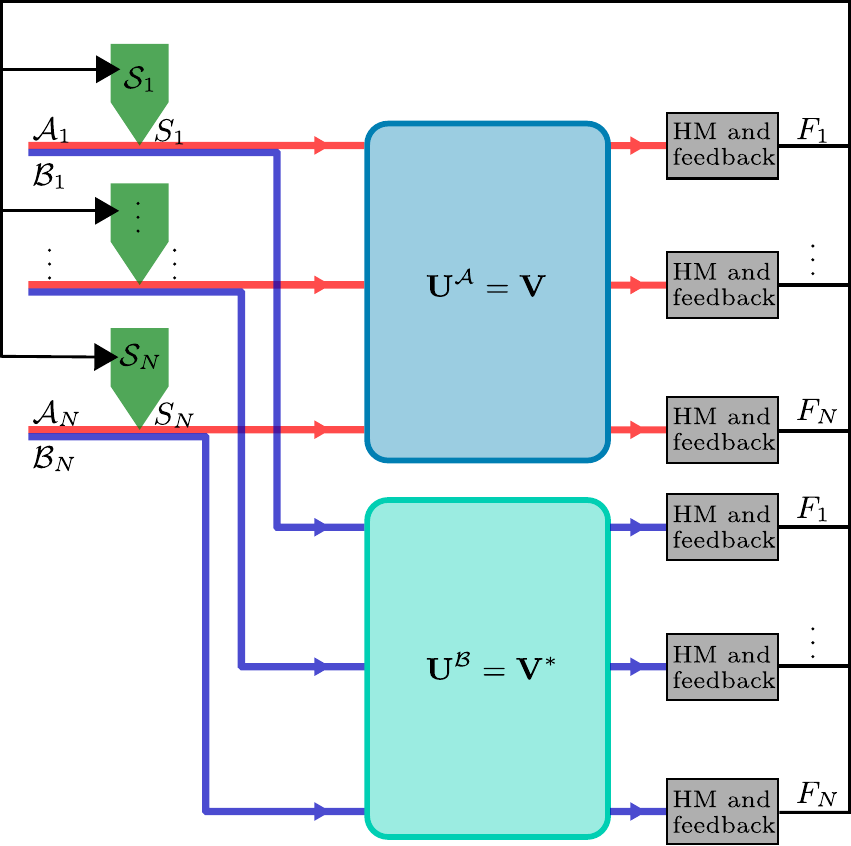}%
	\caption{A feedback scheme to realize an open Ising model. Each two-level system $\sys{S}_{k}$ couples to two light fields, $\sys{A}_{k}$ (red) and $\sys{B}_{k}$ (blue). The $\sys{A}$-fields traverse an interferometer $\mat{V}$, while the $\sys{B}$-fields pass the complex conjugate $\mat{V}^*$. A homodyne measurement (HM) of each field then allows for feedback via operators $\op{F}_{k}$. This way one can realize the dynamics of a dissipative Ising model. \label{fig:QuantumSimulation:Ising}}
\end{figure}

To generate this Hamiltonian using our feedback scheme, we consider an ensemble of $N$ two-level systems, $\sys{S}_{1},\dots,\sys{S}_{N}$, as in the previous section. Each $\sys{S}_{k}$ is coupled to two light fields, $\sys{A}_{k}$ and $\sys{B}_{k}$, as shown in Fig.~\ref{fig:QuantumSimulation:Ising}. We choose identical system-light coupling for both fields via system operators
\begin{align}
	\op{S}_{k}^{\sys{A}} = \op{S}_{k}^{\sys{B}} =: \op{S}_{k} = \hat{\sigma}_{-}^{k},
\end{align}
as well as identical feedback operators
\begin{align}
	\op{F}_{k}^{\sys{A}} & = \op{F}_{k}^{\sys{B}} =: \op{F}_{k} = \sum_{j=1}^{N} g_{jk}\hat{\sigma}_{x}^{j}, \label{eq:QuantumSimulation:Feedback}
\end{align}
with some feedback gains $\mat{G}=(g_{jk})\in\mathbb{R}^{N\times N}$. Note that even though system and feedback operators are the same, the setups may still generate different dynamics if their interferometers $\mat{U}^\sys{A}$ and $\mat{U}^\sys{B}$ are different.

We first neglect the $\sys{B}$-fields, and consider only the dynamics generated by the $\sys{A}$-fields. We fix the interferometer $\mat{U}^{\sys{A}}$ to be some $N\times N$ unitary matrix, $\mat{U}^{\sys{A}}=\mat{V}$. Together with the operators $\op{S}_{k}$ and $\op{F}_{k}$ chosen as above, we find the feedback Hamiltonian
\begin{align}
	\op{H}^{\sys{A}} & = \frac{1}{2}\sum_{k\neq l}  \left( \mrm{Re}[K_{kl}]\hat{\sigma}_{x}^{l} + \mrm{Im}[K_{kl}] \hat{\sigma}_{y}^{l} \right) \hat{\sigma}_x^{k} \notag\\
	& \qquad + \frac{1}{2}\sum_{k=1}^{N} \mrm{Re}[K_{kk}] (\op{1}+\hat{\sigma}_{z}^{k}),
\end{align}
where we defined the matrix $\mat{K}=(K_{kl}):=\mat{G}\mat{V}$, comprising gains and interferometer, and $\mrm{Re}[\cdot]$ and $\mrm{Im}[\cdot]$ denote real and imaginary part respectively. The Hamiltonian $\op{H}^{\sys{A}}$ already resembles $\op{H}_{\mrm{Ising}}$ from Eq.~\eqref{eq:IsingModel:Hamiltonian}, up to terms creating a $\hat{\sigma}_{x} \hat{\sigma}_{y}$-interaction. We use the second feedback setup to eliminate these.

For the moment, let us consider only the dynamics generated by the $\sys{B}$-fields. We choose the second interferometer to be the complex conjugate of the first, $\mat{U}^{\sys{B}}=\mat{V}^{*}$.  Thus we obtain a second feedback Hamiltonian $\op{H}^{\sys{B}}$ with $\mat{K}^*$ instead of $\mat{K}$, so
\begin{align}
	\op{H}^{\sys{B}} & = \frac{1}{2}\sum_{k\neq l}  \left( \mrm{Re}[K_{kl}]\hat{\sigma}_{x}^{l} - \mrm{Im}[K_{kl}] \hat{\sigma}_{y}^{l} \right) \hat{\sigma}_x^{k} \notag\\
	& \qquad + \frac{1}{2}\sum_{k=1}^{N} \mrm{Re}[K_{kk}] (\op{1}+\hat{\sigma}_{z}^{k}).
\end{align}

As explained in Section \ref{sec:Derivation:MultipleLightModes}, we obtain the combined evolution of both schemes by adding their Liouvillians. Thus, the total Hamiltonian $\op{H} = \op{H}^{\sys{A}} + \op{H}^{\sys{B}}$ will be
\begin{align}
	\op{H} & = \sum_{k\neq l} \mrm{Re}[K_{kl}]\hat{\sigma}_x^{k}\hat{\sigma}_{x}^{l} + \sum_{k=1}^{N} \mrm{Re}[K_{kk}] (\op{1}+\hat{\sigma}_{z}^{k}). \label{eq:IsingModel:FeedbackHamiltonian}
\end{align}
The interaction and local fields in this model are entirely governed by the matrix $\mrm{Re}[\mat{K}] =\mrm{Re}[\mat{G}\mat{V}] = \mat{G}\mrm{Re}[\mat{V}]$. Recall that we are free to choose both the feedback gains $\mat{G}$ and the interferometer $\mat{V}$. Assuming $\mrm{Re}[\mat{V}]$ can be inverted, we can realize a given set of couplings $\mat{\Delta}$ by setting the gains as
\begin{align}
	\mat{G}= \mat{\Delta}\mrm{Re}[\mat{V}]^{-1}.
\end{align}
This will yield the desired Hamiltonian from Eq.~\eqref{eq:IsingModel:Hamiltonian},
\begin{align}
	\op{H} & = \sum_{k\neq l} \Delta_{kl}\hat{\sigma}_{x}^{k} \hat{\sigma}_{x}^{l} - \sum_{k=1}^{N}B_{k} (\hat{\sigma}_{z}^{k} + \op{1}), \label{eq:Ising:FeedbackIsingHamiltonian}
\end{align}
up to negligible constants.

A generic unitary $\mat{V}$ is not guaranteed to have a non-singular real part. However, writing
\begin{align}
	\mrm{Re}[\mat{V}] & = \frac{1}{2}(\mat{V}^*+\mat{V}) = \frac{1}{2}\mat{V}^* (\mat{1}_{N}+\mat{V}^\transpose \mat{V})
\end{align}
shows that invertibility of $\mrm{Re}[\mat{V}]$ is equivalent to that of $\mat{1}_{N}+\mat{V}^\transpose \mat{V}$. Thus the necessary and sufficient condition is that all eigenvalues of $\mat{V}^\transpose \mat{V}$ are different from $-1$. Note also, that $\mat{V}$ does not need to reflect the geometry of the physical model as given by $\mat{\Delta}$.

\subsubsection{Effect of the dissipation}
Recall from Eq.~\eqref{eq:Derivation:MultiSysMasterEquation} that our scheme naturally introduces jump operators of the form
\begin{align}
	\op{\tilde{J}}_{k} := \op{Z}_{k}-i\op{F}_{k} = \sum\nolimits_{l}U_{kl}\op{S}_{l}-i\op{F}_{k}
\end{align}
for each light field. In this case, there will be two sets of jump operators, corresponding to the $\sys{A}$- and $\sys{B}$-fields respectively.

As mentioned in Section \ref{sec:LOCC}, a collection of jump operators $\op{J}_{k}$ is always invariant under unitary transformations,
\begin{align}
	\sum_{k}\mcl{D}[\op{J}_{k}]\rho & = \sum_{k}\mcl{D}\Big[\sum\nolimits_{l}U_{kl}\op{J}_{l}\Big]\rho.
\end{align}
We can use this to define a new set of jump operators,
\begin{align}
	\op{J}_{k} := \sum_{l=1}^{N}U_{lk}^*\op{\tilde{J}}_{l} = \op{S}_{k}-i\sum_{l=1}^{N}U_{lk}^{*} \op{F}_{l},
\end{align}
equivalent to the former. These are given by
\begin{align}
	J_k^{\sys{A}} & :=\op{S}_{k}-i\sum_{j=1}^{N}V_{jk}^* \op{F}_{j} = \hat{\sigma}_{-}^{k} - i \sum_{j=1}^{N} \Gamma_{jk} \hat{\sigma}_x^{j},\\
	J_k^{\sys{B}} & :=\op{S}_{k}-i\sum_{j=1}^{N}V_{jk} \op{F}_{j} = \hat{\sigma}_{-}^{k} - i \sum_{j=1}^{N} \Gamma_{jk}^{*} \hat{\sigma}_x^{j},
\end{align}
with the complex matrix
\begin{align}
	\mat{\Gamma} & := \mat{\Delta} \mrm{Re}[\mat{V}]^{-1} \mat{V}^* = 2\mat{\Delta}(\mat{1}_N+\mat{V}^\transpose \mat{V})^{-1}.
\end{align}

Combining the Hamiltonian and jump operators yields the dynamics of an open system of interacting spins. The final master equation is
\begin{align}
	\dot{\rho} & = -i[\op{H},\rho] + \sum_{k=1}^{N}\mcl{D}[\op{J}_{k}^{\sys{A}}]\rho + \sum_{k=1}^{N}\mcl{D}[\op{J}_{k}^{\sys{B}}]\rho,
\end{align}
with the Hamiltonian from Eq.~\eqref{eq:Ising:FeedbackIsingHamiltonian}, and jump operators as above.

An important observation is that the coupling matrix $\mat{\Delta}$ appears in $\mat{\Gamma}$ and thus also in the jump operators. Hence, fixing the interaction and local fields in the Hamiltonian also determines the structure and strength of the dissipation. In particular, both $\mat{\Delta}$ and $\mat{V}$ will determine the range of the jump operators, i.e., how many systems couple to the same bath.

Note that there are degrees of freedom we have not used. One parameter we could change is the relative strength of the system-light interaction and feedback. We can change the interaction by a factor $r\in\mathbb{R}$, i.e., $\op{S}\mapsto r \op{S}$, while inversely changing the feedback, $\op{F}\mapsto \op{F}/r$. This will not affect the Hamiltonian since it only comprises products of the form $\op{S}\op{F}$. On the other hand, it will cause either $\hat{\sigma}_{-}$ (for $r\gg 1$) or $\hat{\sigma}_{x}$ (for $r\ll 1$) to dominate the jump operators, and also increase the overall strength of the dissipation compared to $\op{H}$.

Another free parameter is asymmetry of $\mat{\Delta}$. Let us set $\mat{\Delta}_{\pm}:=(\mat{\Delta}\pm \mat{\Delta}^\transpose)/2$ as the symmetric and skew-symmetric part of $\mat{\Delta}=\mat{\Delta}_{+}+\mat{\Delta}_{-}$ respectively. The symmetric part is fixed by the physical coupling in the Hamiltonian from Eq.~\eqref{eq:IsingModel:Hamiltonian}. However, $\mat{\Delta}_{-}$ does not enter the Hamiltonian, so we are free to choose it at will. It only affects the jump operators. For simplicity, we set $\mat{\Delta}_{-} = \mbf{0}_{N}$ in the following example.

\subsection{Concrete example\label{sec:Ising:NearestNeighborInteraction}}
To demonstrate how to realize a specific model, we consider a translationally invariant one-dimensional Ising chain with nearest-neighbor interaction and periodic boundary conditions. The corresponding physical Hamiltonian is
\begin{align}
	\op{H}_{\text{Ising}} & = \Delta\sum_{k=1}^{N} \hat{\sigma}_{x}^{k} \hat{\sigma}_{x}^{k+1} - B\sum_{j=1}^{N} \hat{\sigma}_{z}^{j}, \label{eq:IsingModel:ConcreteHamiltonian}
\end{align}
with periodicity implemented via $\hat{\sigma}_{x}^{N+1}\equiv \hat{\sigma}_{x}^{1}$. We can read off the interaction matrix $\mat{\Delta}$,
\begin{align}
	\mat{\Delta} & = B\mat{1}_{N} + \frac{\Delta}{2}\left( \mat{S}+ \mat{S}^\transpose \right),
\end{align}
where $\mat{S}=(s_{kl})$ is the periodic shift matrix with elements $s_{N,1} = s_{k,k+1}=1\ \forall\, k$ and $s_{kl}=0$ otherwise.

We would now like to engineer translationally invariant jump operators in order not to break the symmetry of the model. To this end we must choose $\mat{V}$ such that the coefficient matrix from the jump operators, $\mat{\Gamma}=2\mat{\Delta}(\mat{1}_{N}+\mat{V}^\transpose \mat{V})^{-1}$, is also translationally invariant. This is the case if and only if $[\mat{V}^\transpose \mat{V},\mat{S}] = 0$, so $\mat{V}^\transpose \mat{V}$ and $\mat{S}$ must be simultaneously diagonalizable. We know that the discrete Fourier transform $\mat{F}=(F_{jk})$ \cite{Idel2015} with
\begin{align}
	F_{jk} & = \frac{1}{\sqrt{N}}\expo{2\pi i\frac{jk}{N}},\qquad  j,k = 0,\dots,N-1,
\end{align}
diagonalizes the shift matrix with $\mat{S} = \mat{F}\mat{\Omega} \mat{F}^\dag$, where $\mat{\Omega} = \diag{\omega_0,\dots,\omega_{N-1}}$ and $\omega_k = \expo{2\pi i \frac{k}{N}}$. Thus we must choose $\mat{V}$ such that
\begin{align}
	\mat{V}^\transpose \mat{V} & = \mat{F}\mat{\Lambda}^2 \mat{F}^\dag
\end{align}
is satisfied for some diagonal unitary matrix $\mat{\Lambda}=\diag{\lambda_0,\lambda_1,\dots,\lambda_{N-1}}$. The combination $\mat{V}^\transpose \mat{V}$ is symmetric, which requires
\begin{align}
	\lambda_{k}=\lambda_{N-k}
\end{align}
for all $k$. A possible solution is to set
\begin{align}
	\mat{V} = \mat{F}\mat{\Lambda}\mat{F}^\dag,
\end{align}
which is also symmetric because $\mat{\Lambda}^2$ and $\mat{\Lambda}$ have the same structure. This leads to the correct expression $\mat{V}^\transpose \mat{V} = \mat{V}^2 = \mat{F}\mat{\Lambda}^2 \mat{F}^\dag$. Since $\mat{1}_{N}+\mat{V}^\transpose \mat{V}$ needs to be non-singular we require that $\lambda_k\neq \pm i$ for all $\lambda_k$.

The independent eigenvalues $\lambda_{k}$ can still be used to change the structure and range of the jump operators. We set
\begin{align}
	\lambda_k^2 = \frac{1-2i\cos(2\pi k /N)}{1+2i\cos(2\pi k /N)},	
\end{align}
for $k=0,\dots,N-1$, to obtain a tridiagonal matrix
\begin{align}
	(\mat{1}_{N}+\mat{V}^\transpose \mat{V})^{-1} = (1/2)(\mat{1}_{N}+i\mat{S}+i\mat{S}^\transpose).
\end{align}
Recall that $\mat{\Delta} = B\mat{1}_{N} + (\Delta/2)\left( \mat{S}+ \mat{S}^\transpose \right)$ is also tridiagonal, so we find that
\begin{align}
	\mat{\Gamma} & = (B + i\Delta)\mat{1}_{N} + (\frac{\Delta}{2}+iB)(\mat{S}+\mat{S}^\transpose)\notag\\
	&\quad + i\frac{\Delta}{2}(\mat{S}^2 + (\mat{S}^{\transpose})^{2}).
\end{align}
This matrix couples only nearest and next-nearest neighbors, so each jump operator $\op{J}_{k}$ will act on the 5 systems $\sys{S}_{k-2},\dots,\sys{S}_{k+2}$ for any $k$. Nearest-neighbor-only jump operators would require a diagonal $\mat{V}^\transpose \mat{V}$, which in turn implies $\mat{V}=\mat{O\Lambda}$ for some real orthogonal matrix $\mat{O}$ and diagonal unitary matrix $\mat{\Lambda}$. This, however, places us in the LOCC regime discussed in Section \ref{sec:LOCC}, which would erase possible quantum features from the model.

In addition to restricting the range of the jump operators, we can change the relative strength of system and feedback operators, as mentioned in the previous section. This will leave $\op{H}$ invariant, but introduce a parameter $r\in\mathbb{R}$ in the jump operators,
\begin{align}
	\begin{split}
		J_k^{\sys{A}} & = r\hat{\sigma}_{-}^{k} - \frac{i}{r}\sum_{j=1}^{N} \Gamma_{jk} \hat{\sigma}_x^{j},\\
		J_k^{\sys{B}} & = r\hat{\sigma}_{-}^{k} - \frac{i}{r}\sum_{j=1}^{N} \Gamma_{jk}^* \hat{\sigma}_x^{j}.
	\end{split}
\end{align}

For large $r\gg 1$, the dissipation will be dominated by local $\hat{\sigma}_{-}$-decay, driving each spin into the $|\downarrow_{z}\rangle$-state. In the limit of small $r\ll 1$, the $\hat{\sigma}_{x}$-dephasing takes over. Note that the sum
\begin{align}
	\op{\Sigma}_{x}^k & := \sum_{j=1}^{N} \Gamma_{jk}^* \hat{\sigma}_x^{j}
\end{align}
may in principle generate interesting quantum states. For instance, when applied to the trivial state $\bigotimes\nolimits_{j}|\downarrow_{z}\rangle_{j}$ it will create a kind of $|W\rangle$ state with weights $\Gamma_{jk}$. On the other hand, in the case of strong dissipation governed by $\op{\Sigma}_{x}$, the systems will likely end up in a mixed state, since $[\op{\Sigma}_{x},\op{\Sigma}_{x}^{\dag}]=0$, so $\mcl{D}[\op{\Sigma}_{x}]\mbf{1}=0.$

\begin{figure}
	\includegraphics[width=\columnwidth,keepaspectratio=true]{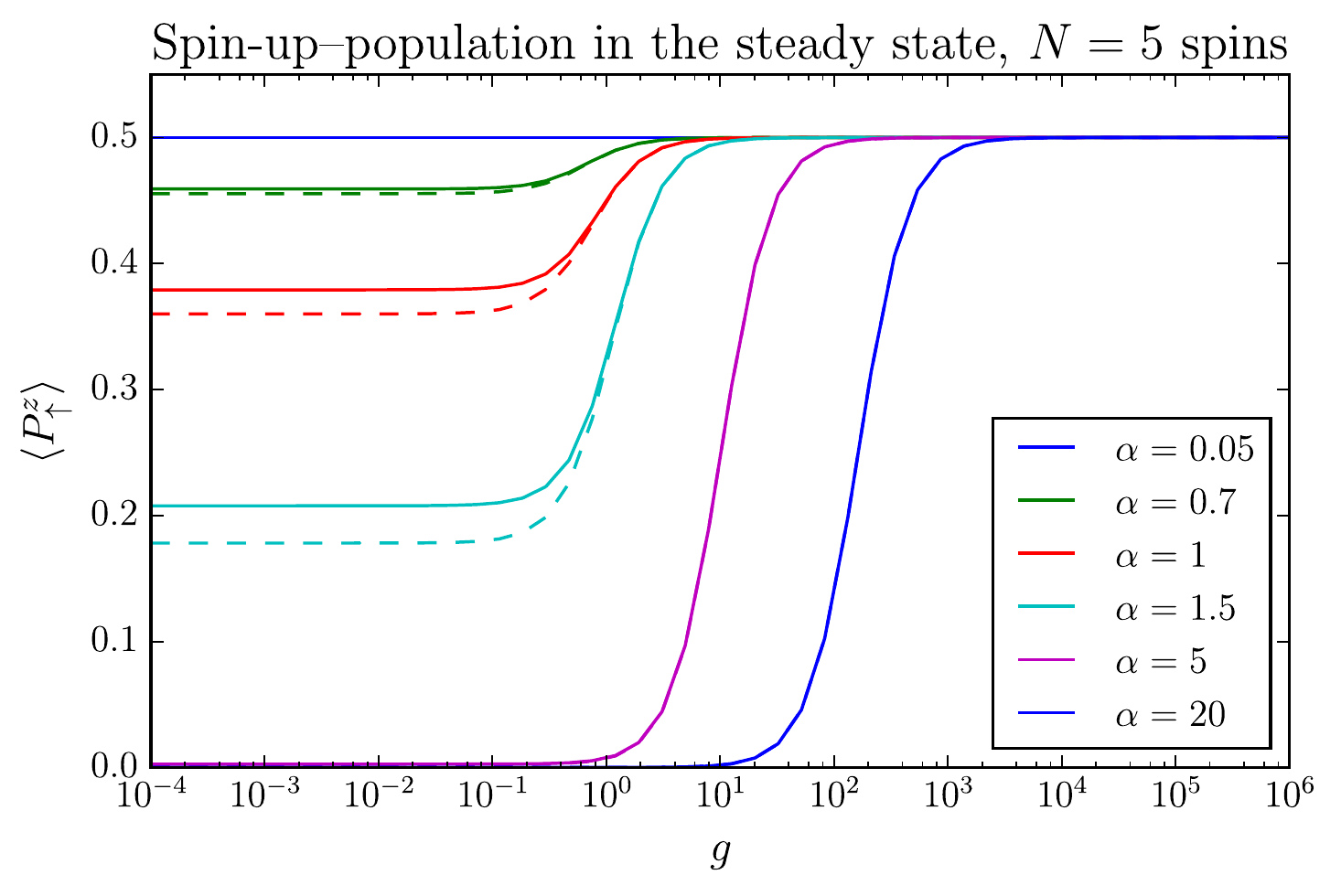}%
	\caption{We plot the spin-up--density in the steady state of the dissipative Ising model against $g=B/\Delta$ (continuous lines) for $N=5$ spins. We consider different values of the asymmetry parameter $\alpha=r/|\Delta|^{1/2}$. For comparison we also plot the results of the purely dissipative steady states obtained by setting $\op{H}=0$ in the dynamics (dotted lines). A small deviation between the full and purely dissipative steady states occurs only when $\alpha$ is close to unity. \label{fig:QuantumSimulation:Plots}}
\end{figure}

\subsubsection*{Results}
We set $g:=B/\Delta$ and $\alpha:=r/|\Delta|^{1/2}$, and examine the steady state of the master equation for different values of $g$ and $\alpha$ at unit interaction strength $\Delta=1$. We can distinguish the following limiting cases.

When $g\ll 1$, the Hamiltonian $\op{H}$ will be governed by the interaction, $\op{H}\sim\sum\nolimits_{k}\hat{\sigma}_{x}^{k}\hat{\sigma}_{x}^{k+1}$. We find that the dominant elements of $\mat{\Gamma}$ are of order $\mcl{O}(1)$ when $g\ll 1$, so the dissipation only depends on our choice of $\alpha$. For $\alpha\gg 1$, the local $\hat{\sigma}_{-}$-decay will dominate the master equation, and all spins align in the $|\downarrow_{z}\rangle$-state. For $\alpha\ll 1$ we expect the $\op{\Sigma}_{x}$-dephasing to generate a highly mixed steady state.

In the limit of $g\gg 1$ and $\alpha\ll 1$, the main contribution to the Hamiltonian will come from the local fields proportional to $g$, so $\op{H}\sim -g\sum\nolimits_{k}\hat{\sigma}_{z}^{k}$. However, the total master equation will again be dominated by $\op{\Sigma}_{x}$-dephasing, creating a mixed steady state. The reason is that the dominant elements of $\mat{\Gamma}$ are now also of order $\mcl{O}(g)$, which enhances the dissipation rate by $g^2/\alpha^2\gg g$.

For $g,\alpha\gg 1$ we can distinguish two different cases. When $\alpha^2 \gg g \gg 1$, the jump operators are dominated by $\alpha\hat{\sigma}_{-}^{k}$, since the largest elements of $\mat{\Gamma}$ are of order $\mcl{O}(g)$ so the $\op{\Sigma}_{x}$-dephasing goes as $g/\alpha\ll \alpha$. Conversely, for $g\gg \alpha^2 \gg 1$ the dephasing takes over and creates a mixed steady state.

Notably, in both limiting cases, $\alpha\gg 1$ and $\alpha\ll 1$, the dissipation dominates the dynamics, independent of $g$. We expect the Hamiltonian to play a role only when $\alpha$ is close to unity. To see this effect, we computed the steady state using QuTiP \cite{Johansson2012,Johansson2013a} for $N=5$ spins. We chose the density of spins in the up-state $|\uparrow_{z}\rangle$,
\begin{align}
	\langle \op{P}_{\uparrow}^{z} \rangle := \frac{1}{N}\sum_{j=1}^{N}\frac{1+\langle \hat{\sigma}_{z}^{j} \rangle}{2},
\end{align}
as an order parameter.

The results of the full dynamics are shown as solid lines in Fig.~\ref{fig:QuantumSimulation:Plots}. They behave as expected in the limits we just discussed. The plots also illustrate the crossover in the intermediate regime for $\alpha\approx 1$ and $g\ll 1$, where $\hat{\sigma}_{-}$ and $\op{\Sigma}_{x}$ compete. To isolate the effect of the Hamiltonian, we also computed the steady state of the purely dissipative dynamics, setting $\op{H}=0$ (dotted lines in Fig.~\ref{fig:QuantumSimulation:Plots}). As expected, the dissipation is dominant whenever $\alpha$ is different from unity, and there is only a slight deviation for $\alpha\approx 1$.

In the regime $g, \alpha \gg 1$, the systems behave qualitatively similar to a dissipative Rydberg gas \cite{Hu2013,Ates2012,Lee2011,Marcuzzi2014,Hoening2014}. This is not entirely surprising: in both cases, the dissipation breaks the $\mathbb{Z}_2$ symmetry of the Ising Hamiltonian, hinting that both models could fall into the same universality class. The dissipative Rydberg gas exhibits a liquid-gas transition for sufficiently strong driving \cite{Weimer2015}, corresponding to the steep increase in Fig.~\ref{fig:QuantumSimulation:Plots}. However, as the liquid-gas transition belongs to the universality class of the classical Ising model \cite{Marcuzzi2014}, we cannot expect to observe a true phase transition in our 1D simulations.

\section{Conclusions and outlook\label{sec:Conclusion}}
We considered setups composed of arrays of light-matter interfaces emitting continuous-wave light which is mixed in a linear optical interferometer and measured in continuous homodyne detections. The corresponding photocurrents are used to perform continuous-time Markovian feedback on the material systems. We presented an elementary derivation of the corresponding feedback master equation, and found a sufficient condition for the scheme to generate only LOCC dynamics precluding the occurrence of genuine quantum effects. We investigated whether the scheme could be used to produce entangled many-body states, and confirmed this by designing a specific setup whose unique steady state has non-zero concurrence and logarithmic negativity. Lastly, we showed that it is possible to engineer the dynamics of an open Ising model. 

The aim of this paper was to identify the general class of quantum dynamics encompassed by the given setup. This turned out to be surprisingly rich, which, as we hope, became clear in this article.
The specific realizations studied in Sections \ref{sec:StatePreparation} and \ref{sec:QuantumSimulation} %clearly
demonstrate that the general FME comprises nontrivial dynamics for appropriate choices of system and feedback operators.

Since our main goal was to gauge the scope of the feedback setup, a number of interesting questions was left unanswered. For example, is it possible to classify, at least partially, the set of reachable steady states and realizable physical models? Can one develop a systematic procedure to find operators and interferometer which prepare a given quantum state?
The FME was derived under ideal conditions, which can only be recovered in the limit of large cooperativity of the light-matter interface. Thus, for specific applications the general FME has to be appended with a more complete description covering finite efficiencies, competing decoherence, and feedback delays which may also enrich the dynamics if chosen carefully.

We hope that our work might trigger further investigations leading to a more comprehensive understanding of the general %model
framework developed here.
In view of the tremendous experimental progress with light-matter interfaces and continuous feedback operations we believe that our approach offers interesting perspectives for future developments in this direction.

%In view of the tremendous experimental progress with light-matter interfaces and continuous feedback operations we believe that our work offers interesting perspectives for future developments in this direction. It is clear that for specific applications the general FME has to be appended with more complete descriptions covering finite efficiencies, competing decoherence, feedback delays etc. Our main aim here was rather to identify the general class of quantum dynamics encompassed by the given setup which indeed is surprisingly rich which, as we hope, became clear in this article.

\section{Acknowledgments}
We thank Sebastian Hofer for discussions. We acknowledge support from DFG through SFB 1227 (DQ-mat), from the European Commission through FET-Open Project iQOEMS, and from the Volkswagen Foundation.

%\appendix
%\input{Chapters/Appendix}

%\bibliography{library} % library.bib is my whole library. Use JabRef to create the reducedlibrary.bib file only from the actually cited references.
\bibliography{reducedlibrary}

\end{document}